\documentclass[3p,times,numbers,sort]{elsarticle} %\documentclass[review]{elsarticle} 
\usepackage{lineno,hyperref}
\usepackage{amsmath,amsfonts,amsthm,graphicx, commands}
\usepackage{enumitem}
\usepackage{breqn}
\usepackage{algorithm}
\usepackage{algpseudocode}
\usepackage{setspace}
\usepackage{mathtools}
\usepackage{subcaption}
\usepackage{ragged2e} 
\usepackage{fancyhdr}
\usepackage{natbib}
\usepackage[dvipsnames]{xcolor}

\journal{Journal of Renewable Energy}

\begin{document}

\begin{frontmatter}

\title{Strongly-coupled aeroelastic free-vortex wake framework for floating offshore wind turbine rotors. Part 2: Application}

\author[]{Steven N. Rodriguez\fnref{fn1} \corref{cor1}}
\fntext[fn1]{snr214@lehigh.edu}\cortext[cor1]{Corresponding author}

\author[]{Justin W. Jaworski\fnref{fn2}}
\fntext[fn2]{jaworski@lehigh.edu}
\address{Department of Mechanical Engineering and Mechanics, Lehigh University, Bethlehem, Pennsylvania 18015-3085, USA }

%\author[mymainaddress,mysecondaryaddress]{Lehigh University}



\begin{abstract}

This two-part paper presents the integration of the free-vortex wake method (FVM) with an aeroelastic framework suitable to model the rotor-wake interactions engendered by floating offshore wind turbine (FOWT) rotors in operation. Part 1 of this paper introduces the numerical development and validation of an aeroelastic framework. Due to a lack of experimental aeroelastic benchmarks for FOWTs, a series of validation studies are conducted against the rotor aerodynamic and structural performance of the National Renewable Energy Laboratory (NREL) 5-MW reference wind turbine. Part 2 of this paper focuses on the modeling and simulating different aeroelastic operational conditions of FOWTs. Numerical results of the current framework capture consistently the aerodynamic rotor performance, such as power, thrust, and torque of wave-induced pitching FOWTs. In addition, the presented aeroelastic framework yields additional information about the power, thrust, and torque fluctuations due to the out-of-phase blade passing frequency and corresponding blade deflections. The fidelity of the presented framework demonstrates, for the first time, an FVM-based aeroelastic method capable of carrying out investigations on rotor-wake interactions and relevant aeroelastic phenomena of FOWTs.
\end{abstract}


\end{frontmatter}

%\linenumbers

\section{Introduction}
Modeling the aerodynamic-elastic (aeroelastic) behavior of rotors is essential to comprehend complex rotor-wake interactions (RWIs) of floating offshore wind turbines (FOWTs) caused by wave-induced motions, which move FOWT rotors into and away from their wake about six degrees of freedom. However, there is a lack of adequate computational frameworks that can capture the complexities of FOWT RWI's at computational costs acceptable to conduct engineering analysis and design. Recently Liu \cite{liu2018aeroelastic} has presented the coupling of the free-vortex wake method for FOWTs, WInDS, developed by Sebastian and Lackner \cite{snl:5, sebastian:10} and later improved by Gaertner and Lackner, \cite{gaertner2015modeling, gaertner2016improved}, to FAST,  an aero-servo-elastic numerical framework developed by the National Renewable Energy Laboratory (NREL) \cite{jonkman2005fast}. The investigation conducted by Liu \cite{liu2018aeroelastic} presents the fluid-structure interaction as one-way coupled, whereby WInDS sends aerodynamic loads to FAST without an iterative feedback loop to converge the aerodynamic (WInDS) and structural (FAST) modules at each time step. Liu \cite{liu2018aeroelastic} also integrates a tree-code algorithm to accelerate the $N$-body computations posed by the free-vortex wake method. However, to the best of the authors' knowledge, the aeroelastic framework presented herein is the first self-contained and strongly-coupled aeroelastic framework in the literature that can adequately capture the complexities of FOWT RWIs at computational costs that are acceptable to conduct engineering analysis and design. Part 1 of this two-part paper \cite{rodriguez_JRE_p1} addressed the lack of an adequate aeroelastic framework capable of capturing FOWT RWIs by presenting a strongly-coupled aeroelastic free-vortex wake framework, which was validated against the fixed tower NREL 5-MW reference wind turbine. The purpose of the present work, Part 2, is to demonstrate the application of this self-contained strongly-coupled aeroelastic framework to simulate and investigate the physics and performance of relevant FOWT configurations by modeling, namely the offshore NREL 5-MW reference wind turbine at below-rated conditions atop a spar-buoy floating platform, and at rated and above-rated conditions atop a barge floating platform.

It is important to state that although the aeroelastic framework presented in \cite{rodriguez_JRE_p1} was validated against the aeroelastic performance of a fixed-tower NREL 5-MW reference turbine, the presented framework cannot be validated explicitly for FOWT configurations because benchmark data do not yet exist that properly account for RWIs and blade flexibility. For instance, experimental tests on a scaled-down OC4-DeepCWind semi-submersible FOWT are presented in \cite{koch2016validation}, but the physical models of these experiments are scaled down in size and are designed to maintain similitude with respect to the Froude number. The Froude number, a dimensionless parameter that defines the ratio between inertial forces to gravitational forces in a fluid \cite{bredmose2012and}, is the most-often conserved dimensionless parameter for wind-wave basin experiments that helps ensure hydrodynamic similitude between the full and model scales \cite{bredmose2012and}. However, FOWT experimental models cannot attain both Froude and Reynolds number similitude in existing wind-wave testing facilities, as is possible for helicopter testing; full similitude would require a change in fluid viscosity or density \cite{bredmose2012and}. To achieve similitude in aerodynamic blade loading and to scale down the rotor thrust properly between the full-scale and model-scale FOWT, specially-designed rotor blades were implemented into the experimental setup, where these designed rotor blades were nearly rigid \cite{koch2016validation}. The resulting rotor blade design was appreciably different than the full scale rotor blades, in which blade dynamics validation would not be adequate due to vastly different frequency content during rotor operation. Another limitation encountered in using Froude-scaled FOWT experiments to validate aeroelastic behavior is that these smaller-size experimental models cannot maintain the correct aerodynamic behavior of the full-scale FOWT \cite{koch2016validation}. Therefore, the wake dynamics of the scaled-down experiments cannot validate the anticipated RWIs during FOWT operation. Other experimental tests on the OC4-DeepCWind semi-submersible FOWT exist, but these experiments also scaled down their experiment designs with fixed Froude number, where the rotor blades were also designed to be nearly-rigid to eliminate the possibility of significant aeroelastic interactions \cite{coulling2013validation, goupee2012model,goupee2014experimental}.

In addition to a lack of FOWT experimental data to validate the presented framework, it was also discussed in Part 1 of this paper that there is a lack of adequate aeroelastic numerical studies that could be employed as benchmarks to test the presented framework for FOWTs \cite{rodriguez_JRE_p1}. Thus, the purpose of Part 2 of this paper is to present aeroelastic data that could serve as an initial numerical benchmark study of FOWTs. However, to reinforce the presented framework and bring a degree of confidence to the aeroelastic modeling, the results presented are compared against rigid-rotor FOWT simulations that employ the Wake Induced Dynamic Simulator (WInDS), a free-vortex wake aerodynamic framework developed by Sebastian and Lackner \cite{snl:5, sebastian:10}.

The remainder of this paper is organized as follows. Section \ref{FOWTs} introduces different floating platform concepts of the NREL 5-MW floating offshore wind turbine to provide context for the floating platform concepts used in the aeroelastic simulations. Section \ref{aeroelastic} provides a brief overview of the aeroelastic numerical framework used to simulate the FOWTs. Section \ref{Prescribed_Motion} describes the implementation of prescribed wave-induced motions of FOWT rotors into the aeroelastic framework presented in Part 1 of this paper. Section \ref{FOWT_results} presents the resulting rotor aerodynamic loading, structural dynamics, and rotor performance across below-rated, rated, and above-rated results. Finally, Section \ref{Conclusions} presents concluding remarks from this work.

\section {Concepts of the NREL 5-MW Floating Offshore Wind Turbine} \label{FOWTs}
\begin{figure}[h!]
    \centering
    \includegraphics[scale=0.65,trim=1cm 9.5cm 0 9.75cm]{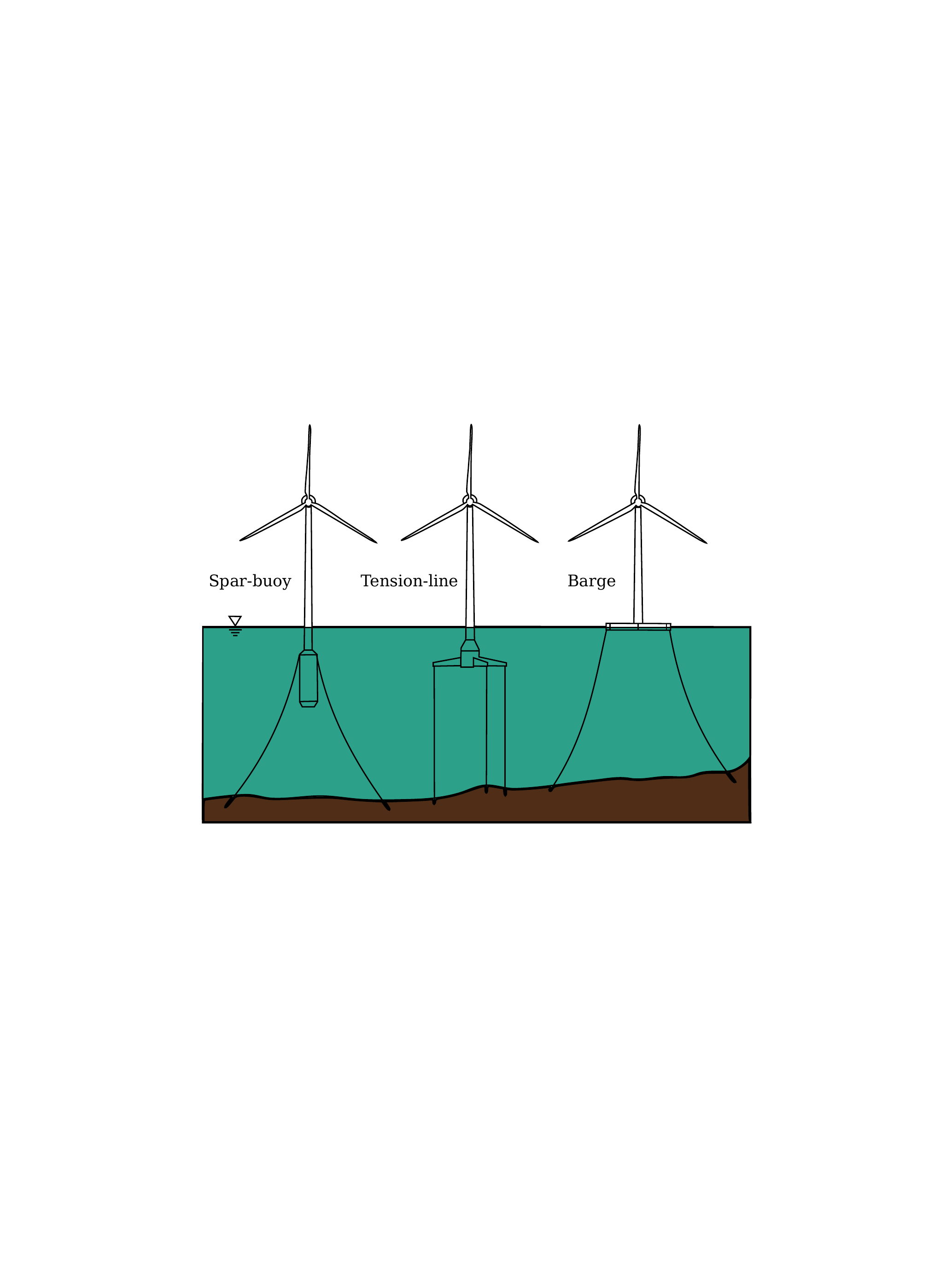}
    \caption[Floating offshore wind turbine concepts]{Floating offshore wind turbine concepts: spar buoy (left), tension-line platform (middle), and barge (right). Adapted from \cite{butterfield2007engineering}.}
    \label{NREL_sim_platforms}
\end{figure}

A brief background is now  provided on a few floating platform conceptual designs that are meant to stabilize  offshore wind turbines that are also emulated in the present numerical framework. The designs chosen for this investigation are described in \cite{jonkman2010offshore, musial:1, musial2016} and are illustrated in Fig.~\ref{NREL_sim_platforms}. The floating platform presented on the left is a spar-buoy concept where its center of mass is below sea-level, which helps stabilize the wind turbine during operation. One can anticipate that the spar-buoy will exhibit large rotor motions for large-amplitude wave forcing with high frequency content.  The middle platform is a tension-line FOWT, where taut lines embedded into the bedrock of the ocean keep the wind turbine stable. One can anticipate that this platform would provide the most stable configuration. However, considerable tension-line fatigue may occur during above-rated conditions that force periodic motion on the wind turbine. Finally, the right platform is a barge platform concept, which employs a wide base to balance the turbine. This barge platform operates much like an oil rig and is attached to a mooring-line system to keep the turbine on station. This barge concept has recently been adapted into a semi-submersible platform concept in the FOWT community, as presented by the OC4-DeepCWind experiments in \cite{koch2016validation}.

\section{Aeroelastic Free-vortex Wake Method}\label{aeroelastic}
To model the FOWT undergoing below-rated, rated, and above rated operational conditions, the aeroelastic framework presented in Part 1 of this paper \cite{rodriguez_JRE_p1} is employed. The aeroelastic framework is based on the lifting line free-vortex wake method strongly coupled to the classical beam theory for spinning structures. A brief review of the underlying aerodynamic and structural theory, their numerical implementation, and the fluid-structural interaction coupling scheme is now presented.

\subsection{Aerodynamics}
The lifting free-vortex wake method assumes that the predominant aerodynamic features in the wake of FOWTs can be modeled by inviscid potential flow theory. This assumption allows the aerodynamic field to be described by the linear superposition of elementary flows to model the freestream velocity and the filament structures to capture the induced velocity of the wake. The movement of the induced velocity field is determined by tracking the discrete filament endpoints, or Lagrangian markers, with position vector $\mathbf{r}$. The velocity of the Lagrangian markers may be written as
\begin{equation}
\frac{d\mathbf{r}}{dt}=\mathbf{U}_{\rm{pl}}+\mathbf{U}_{\rm{induced},\it{i}}+\mathbf{U}_{\infty},
\label{marker_gov_eq1}
\end{equation} 
where $\mathbf{U}_{\rm{pl}}$ consists of the velocity imposed on the FOWT by the offshore environment, $\mathbf{U}_{\rm{induced},\it{i}}$ is the induced velocity of the $i^{\rm{th}}$ filament structure on the Lagrangian marker, and $\mathbf{U}_{\infty}$ is the inflow velocity \cite{snl:5, sebastian:10, rodriguezphd, rodriguezJERT, rodriguez_JRE_p1}. Note that $\mathbf{U}_{\rm{pl}}$ and $\mathbf{U}_{\infty}$ are uniform flow profiles. FVM computes the induced velocity of each $i^{\rm{th}}$ straight segment filament in the wake in a local Cartesian reference frame by the desingularized Biot-Savart law. The current investigation employs the Lamb-Oseen vortex model \cite{lamb1993hydrodynamics} ($n=2$ in Part 1 \cite{rodriguez_JRE_p1}), as experiments suggest that empirically-modified Lamb-Oseen vortex models give excellent comparisons with observed velocity fields surrounding rotor tip-vortices \cite{leishman:2, bhagwat2000correlation}. 

To compute blade loads, the free-vortex wake method is used in conjunction with  lifting-line theory, which concentrates the circulation related to forces generated by a lifting body onto a single bound (lifting-line) filament located at the quarter-chord location along the span of the turbine blade. The lifting-line is also used to compute the circulation of trailing and shed vortex filaments.  The trailing filaments are the connections between shed filaments released at separate instances in time, where shed vortex filaments are initially attached to the trailing edge of the blade and are released at a specified frequency. These trailing and shed vortex filaments make up the spatial and temporal features of the rotor wake, and the FVM framework enforces Kelvin's theorem over the entire wake lattice.

The numerical implementation of this FVM theory involves solving Eq.~\eqref{marker_gov_eq1} via the second order Runge-Kutta scheme and solving the Kutta-Joukowski theorem via fixed-point iteration schemes. The relevant numerical procedures are outlined in more detail in Part 1 of this paper \cite{rodriguez_JRE_p1}.

\subsection{Structural Dynamics}
The classical beam theory for spinning structures \cite{rodriguez_JRE_p1, leung1988spinning} is used to derive the equations of motion for each individual rotor blade. The equations of motion take into account aerodynamic loading and the inertial forces generated by the wave-induced motions of the FOWT rotor. The equations of motion for the rotor blade are as follows.\\

\noindent\emph{\underline{Axial equation of motion}}
\begin{equation}
m\left(2\Omega \dot{v} +\Omega^2\right) + EAu^{\prime \prime} =0, \label{u_eom}
\end{equation}
\emph{\underline{Edgewise equation of motion}}
\begin{equation}
m\left(\ddot{v}+2\Omega\dot{u}-\Omega^2v\right)+ EI_y v^{\prime \prime\prime \prime}-P_{\Omega}v^{\prime\prime}=L_v, \label{v_eom}
\end{equation}
\emph{\underline{Flapwise equation of motion}}
\begin{equation}
m\ddot{w} + EI_z w^{\prime \prime\prime \prime}-P_{\Omega}w^{\prime\prime}=L_w,
\label{w_eom}
\end{equation}
\emph{\underline{Torsional equation of motion}}
\begin{equation}
m\left[\left(I_y+I_z\right) \ddot{\phi} + \Omega^2(I_y-I_z)\phi\right]-GJ\phi^{\prime \prime}=L_{\phi}.\\
\label{phi_eom}
\end{equation}

The variables $u$, $v$, $w$, and $\phi$ are the axial, edgewise, flapwise, and torsional degrees of freedom, and the primes denote derivatives taken with respect to the spanwise variable, i.e.~$d/dx$. The values $m$, $E$, $G$, $A$, $I_y$, $I_z$, and $J$ are the mass per unit length, elastic and shear moduli, cross-sectional area, edgewise and flapwise inertia, and the polar moment of inertia, respectively. Edgewise aerodynamic loading is taken into account by $L_v$, flapwise aerodynamic loading and inertial loading from wave-induced motions are taken into account by $L_w$, and moment aerodynamic loads are taken into account by $L_{\phi}$. It is important to note that the aeroelastic model employed herein and presented in Part 1 of this paper \cite{rodriguez_JRE_p1} considers the impact of the wave-induced motions in the flapwise degree-of-freedom only. Thus, $L_w=L+L_{\rm{rbm}}$, where $L$ is the aerodynamic lift and $L_{\rm{rbm}}$ is the wave-induced inertial loading, which are described in more detail in Section \ref{Prescribed_Motion}.

The numerical framework implemented to solve the rotor blade equations of motion is the traditional linear Galerkin finite-element approach, as presented in \cite{cook2007concepts}. More details of the employed numerical framework are presented in \cite{rodriguez_JRE_p1, rodriguezphd}. The final discretized equation of motion for the blades is expressed as
\begin{equation}
\mathbf{[M]}\mathbf{\ddot{D}}+\mathbf{[C]}\mathbf{\dot{D}}+\mathbf{[K]}\mathbf{D}=\mathbf{F}, 
\label{global_FE_eq}
\end{equation}
where $\mathbf{D}$, $\dot{\mathbf{D}}$, and $\ddot{\mathbf{D}}$ represent the global vectors, which are the collection of local displacements, velocities, and accelerations. The global matrices $\mathbf{M}$, $\mathbf{C}$, and $\mathbf{K}$, are the traditional mass, damping, and stiffness matrices, as presented in~\cite{ferreira2008matlab,cook2007concepts,bathe2}. The vector, $\mathbf{F}$, represents external forces, such as the aerodynamic or the wave-induced inertial forces of an FOWT, 
\begin{equation}
\mathbf{F}=\mathbf{F}_{\rm{aerodynamic}}+\mathbf{F}_{\rm{inertial}},
\end{equation}
where $\mathbf{F}_{\rm{aerodynamic}}$ and $\mathbf{F}_{\rm{inertial}}$ are determined from the discretizing the forces from Eqs.~\eqref{u_eom} - \eqref{w_eom}. The final discrete equation of motion is integrated in time via the Newmark method presented in \cite{bathe2}.

\subsection{Fluid-Structure Interaction Coupling}

The aeroelastic framework employed is a partitioned framework that solves the aerodynamic and structural equations of motion separately and strongly coupled via the Aitken $\Delta^2$ method  \cite{weghs:10, erbts2014acc}. At each time step, the aerodynamic and structural solutions are iterated to achieve kinematic and dynamic continuity at the fluid-structure interface. To satisfy the kinematic continuity conditions at each time step, the Aitken scheme imposes a relaxation of the fluid (lifting-line) and structural (finite element) solution vectors by computing a relaxation factor that minimizes the solution residuals between subiterations in a least-squares sense of the $L_{2}$-norm \cite{rodriguez_JRE_p1,weghs:10,erbts2014acc}. The relaxation method iterates until convergence of the interface location is achieved. The under-relaxation method iterates until the convergence of the interface location is achieved. Further iteration can be used to satisfy the dynamic continuity condition, such that the fluid forces imposed on the structure have converged. A basic overview of this coupling scheme is as follows: 1) compute the aerodynamic blade loads via the Kutta-Joukowski theorem, 2) load blades to compute blade deflections, 3) update the position of the lifting line on the blade, 4) iterate steps 1-3 until the interface (structural and aerodynamic mesh locations) and the aerodynamic and structural forces have converged.

\section{Prescribed Wave-Induced Motion}\label{Prescribed_Motion}

The platforms shown in Fig.~\ref{NREL_sim_platforms} were chosen for this investigation due to the considerable amount of information on their operation made available by Sebastian \emph{et al}.~\cite{sebastian:10, sebastian2012analysis}. In their work, Sebastian \emph{et al}.~emulated FOWT operational conditions by fitting platform-simulated response data using FAST to a two-peak frequency harmonic function, which for a pitching rotor the rigid-body motion is described, in radians, by 

\begin{equation}\label{synthetic_rbm1}
P_{\rm{rbm}}=\frac{\pi}{180^{\circ}}\left(X_0+A_1\sin{\left(2\pi f_1 t+\phi_1\right)}+A_2\sin{\left(2\pi f_2 t+\phi_2\right)}\right).
\end{equation}
 This rigid-body motion was then imposed on the operating wind turbine rotor to emulate operation in offshore conditions. For more detail the reader is directed to~\cite{sebastian2012analysis, sebastian:10}. However, the works conducted by Sebastian \emph{et al}.~\cite{sebastian:10, sebastian2012analysis} did not consider blade elasticity in their models. Thus, the present investigation extends beyond the works of Sebastian \emph{et al}.~by simulating the aeroelastic behavior of the FOWT rotor.

\begin{figure}
    \centering
    \includegraphics[scale=0.25, trim=0 0 0 0]{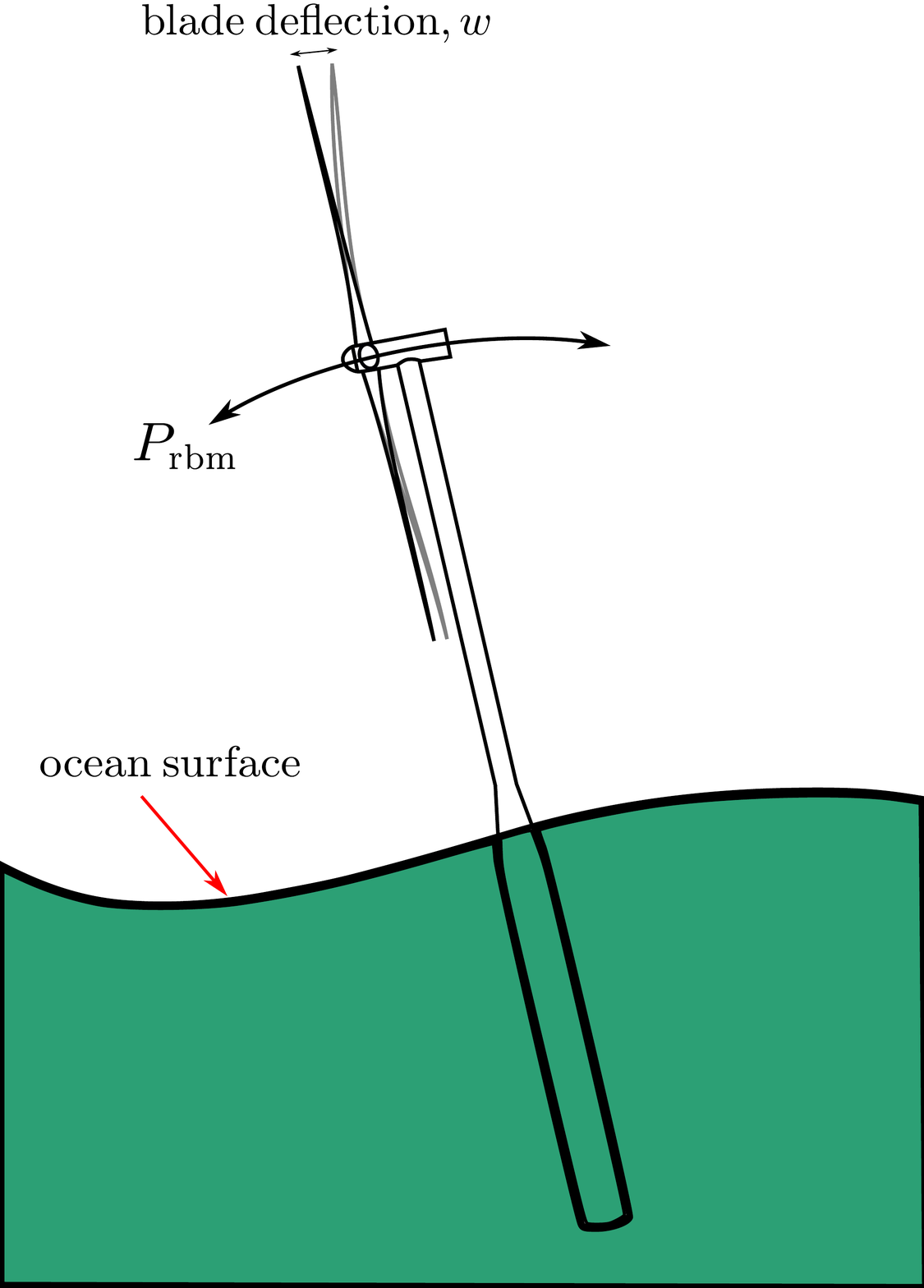}
    \caption{Rigid body motion induced by wave-loading and the offshore environment}
    \label{rbm_inertial}
\end{figure}

The present aeroelastic modeling approach represents the rigid-body motion of the FOWT rotor as an inertial force experienced by the rotor blades in operation. The rigid body motion action is shown in Fig.~\ref{rbm_inertial}. The velocity and acceleration of the pitching motion necessary to describe these inertial forces follow from time derivatives of Eq.~\eqref{synthetic_rbm1}; note that $X_0$ is the initial position of the rotor and is a constant.

To emulate the inertial forces experienced by the rotor blades, the angular acceleration of the rotor is imposed as an unsteady body load on the rotor blades in the blade frame of reference. The inertial loads acting on the rotor blades are assumed to be generated predominantly by the tangential acceleration of the rotor hub, i.e., the product of angular acceleration and the distance to the platform pitching axis. It is further assumed that the platform pitching axis is located at the ocean surface, such that the lever arm is fixed and is approximately the tower height, $h_{\rm{twr}}$. Finally, the pitching acceleration is imposed only as a rectilinear acceleration on the rotor blades, which is equivalent to a small-angle approximation of a pitching FOWT rotor in the ocean environment. The resulting acceleration is an unsteady loading condition across the span of the blade in Eq.~\eqref{w_eom} and is expressed as
\begin{equation}
    L_{\rm{rbm}}=\int_0^R m h_{\rm{twr}}\ddot{P}_{\rm{rbm}} dx,
\end{equation}
where $R$ is the length of the rotor blade and the double overdot denotes a second derivative with respect to time. 

 It is important to note that the treatment of the inertial loading is a crude first approximation of the rotorblade dynamics and rigid-body dynamics. The present work assumes the pitching axis of the FOWT platform structure to be near the surface, which is a valid approximation for barge platforms but is a poorer estimate for spar-buoys whose pitching axes are typically much farther below the ocean surface \cite{robertson2011loads}. With regard to the rotorblade dynamics, as a spanwise section of the rotor blade rotates about the rotor axis, the moment arm between that section and the pitching axis of the FOWT changes in time. Therefore, the inertial loads do not depend solely on the pitching acceleration of the floating platform, but they also depend on the blade rotation frequency; this physical effect has been neglected by the small-angle pitching assumption of the FOWT such that all accelerations of the rotor hub are approximately rectilinear. As mentioned in Part 1 \cite{rodriguez_JRE_p1}, these and other aspects of the current framework motivate further improvement to the modeled physics, specifically with respect to more accurate and faithful representations of rotorblade and rigid-body dynamics by incorporating gyroscopic and Coriolis effects. Nevertheless, the present work is intended to demonstrate the self-contained strongly-coupled framework presented in \cite{rodriguez_JRE_p1} and to establish a foundation onto which additional physical modeling refinements may be integrated.

In the present study, only the OC3-Hywind Spar-Buoy platform is considered at below-rated conditions, and only the ITI Energy Barge platform is considered at rated and above-rated conditions. The respective operational and fitting parameters of the synthetic time series in Eq.~\eqref{synthetic_rbm1}, as defined by \cite{sebastian2012analysis}, are presented in Tables \ref{FOWT_rotoroperation_parameters} and  \ref{FOWT_operation_parameters}. These specific floating platforms at their respective inflow conditions and their prescribe motions were chosen to demonstrate the largest wave-induced rotor motions across the below-rated, rated, above-rated inflow conditions.

{\renewcommand{\arraystretch}{1.5}
\begin{table}[h!]
\centering
\setlength\tabcolsep{3.5pt}
\caption[]{Operational Parameters, where $\theta_{\rm{bl}}$ and $\lambda$ are blade pitch and tip-speed ratio, respectively \cite{sebastian:10, sebastian2012analysis}.}
\begin{tabular}{l c c c c c} 
\hline\hline
Operation & Platform & $V_{\infty}$ (m/s) & $\theta_{\rm{bl}}$ (deg.) & $\lambda$\\
\hline\hline
1.~Below-rated & OC3-Hywind Spar-Bouy & 6 & 0 & 9.63 \\
2.~Rated & ITI Energy Barge & 11.4 & 0 & 7\\
3.~Above-rated & ITI Energy Barge & 18 & 15 & 4.43\\
\hline\hline
\end{tabular}
\label{FOWT_rotoroperation_parameters}
\end{table}}

{\renewcommand{\arraystretch}{1.5}
\begin{table}[h!]
\centering
\setlength\tabcolsep{3.5pt}
\caption[]{FOWT Operational Parameters \cite{sebastian:10, sebastian2012analysis}}
\begin{tabular}{l c c c c c c c c} 
\hline\hline
Operation & $A_1$ & $f_1$ & $p_1$ & $A_2$ & $f_2$ & $p_2$ & $X_0$ & $t_{\rm{lag}}$ \\
\hline\hline
1.~Below-rated&-0.084 & 0.066 & 1.930 & -0.116 & 0.077 & 3.133 & 1.58 & 30\\
2.~Rated & -0.637 & 0.065 & -0.381 & 1.677 & 0.077 & 1.835 & 1.722 & 30\\
3.~Above-rated & 1.518 & 0.066 & 2.132 & 2.979 & 0.078 & 6.863 & 0.939 & 30\\
\hline\hline
\end{tabular}
\label{FOWT_operation_parameters}
\end{table}}

\begin{figure*}[h!]
        \centering
        \begin{subfigure}[b]{0.475\textwidth}
            \centering
            \includegraphics[scale=0.5, trim=160 210 50 250]{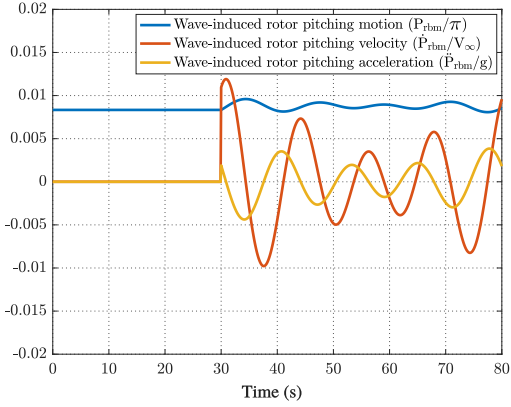}
            \caption[]%
            {{Below-rated wave-induced motions}}    
            \label{BR_rbms}
        \end{subfigure}
        \hfill
        \begin{subfigure}[b]{0.475\textwidth}  
            \centering 
            \includegraphics[scale=0.5, trim=120 210 105 250]{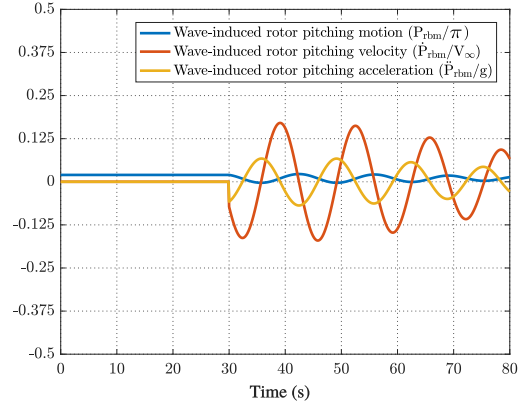}
            \caption[]%
            {{Rated wave-induced motions}}    
            \label{R_rbms}
        \end{subfigure}
        \vskip\baselineskip
        \begin{subfigure}[b]{0.475\textwidth}   
            \centering 
            \includegraphics[scale=0.5, trim=150 210 50 230]{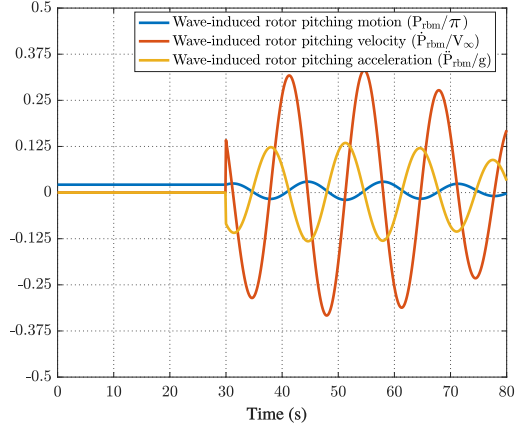}
            \caption[]%
            {{Above-rated wave-induced motions}}    
            \label{AR_rbms}
        \end{subfigure}
        \caption[Time history of wave-induced rigid-body motions]
        {Below-rated, rated, and above-rated time histories of wave-induced motions, velocities, and accelerations} 
        \label{FOWT_platform_timehistories}
    \end{figure*}

The emulated pitching motion, velocity, and acceleration caused by the offshore conditions, for the cases considered here, are shown in Fig.~\ref{FOWT_platform_timehistories}, where the rotor pitch displacement, velocity, and acceleration have been scaled down by a factor of $\pi$ for convenience, non-dimensionalized by the freestream velocity and gravitational acceleration, respectively. 

\section{Floating Offshore Wind Turbine Operational Results}\label{FOWT_results}
The aeroelastic simulations of the floating offshore wind turbines are initiated with a uniform inflow and undeformed rotor blades. For the simulations conducted here, we have employed a 30 second time-lag, $t_{\rm{lag}}$, as listed in Table \ref{FOWT_operation_parameters} so that the initial and artificial transients of the simulations are dissipated before starting the rigid-body motions of the rotor.
Based on the numerical time discretization convergence of the aeroelastic framework in the validation results of Part 1 of this paper \cite{rodriguez_JRE_p1}, the offshore simulations were conducted at a vortex shedding frequency of $f=10$ Hz for the below-rated and above-rated case. However, because of the high blade deflections and added forcing from the wave-induced loads, the rated case employed a vortex shedding frequency of $f=12$ Hz.
Discussions on the wakes generated by the floating offshore wind turbines are presented along with the time histories of aerodynamic loads and performance metrics generated.

\subsection{Below-rated Operation}
\begin{figure}[h!]
\center
\includegraphics[scale=0.125, trim=0cm 75cm 30cm 10cm,clip]{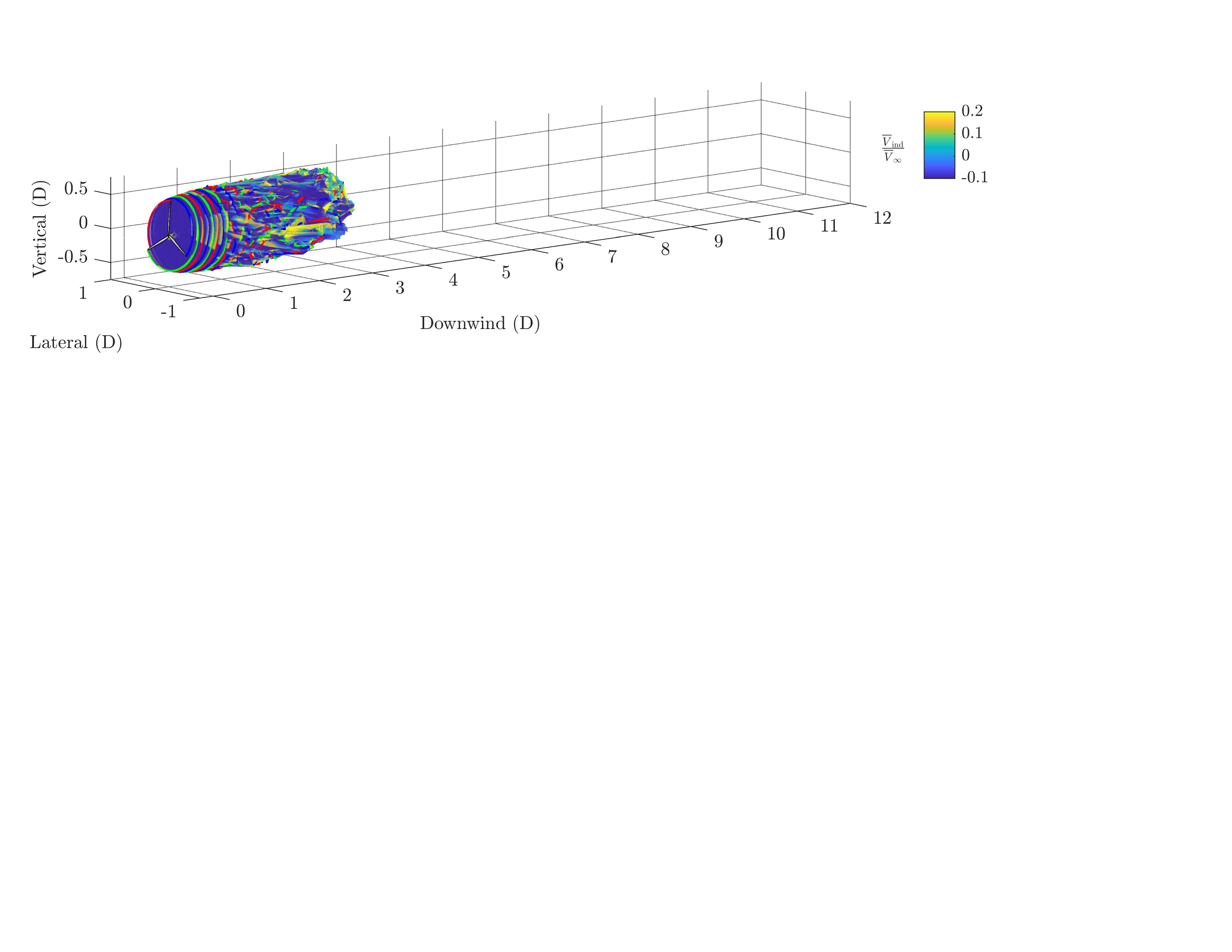}
\caption[FOWT wake shed from a flexible rotor at below-rated operational conditions]{Simulation of the wake shed from a flexible rotor at below-rated operational conditions. The magnitude of the induced velocity and the freestream velocity are $\bar{V}_{\rm ind}$ and $\bar{V}_{\infty}$, respectively. Spatial dimensions are scaled by the rotor diameter, $D$.}
\label{BR_wake_flex_FOWT}
\end{figure}

The wake generated by the below-rated operational conditions of the floating offshore wind turbine is shown in Fig.~\ref{BR_wake_flex_FOWT}. The wake remains coherent for approximately one rotor diameter downstream before the eventual breakdown. The induced velocity opposes the inflow within the first diameter of the downstream region before it begins to break apart. No significant changes appear in the wake formation behind the rotor due to wave-induced forcing of the offshore environment.

\begin{figure*}[p!]
        \centering
        \begin{subfigure}[b]{0.475\textwidth}
            \centering
            \includegraphics[scale=0.225, trim=120 0 50 0]{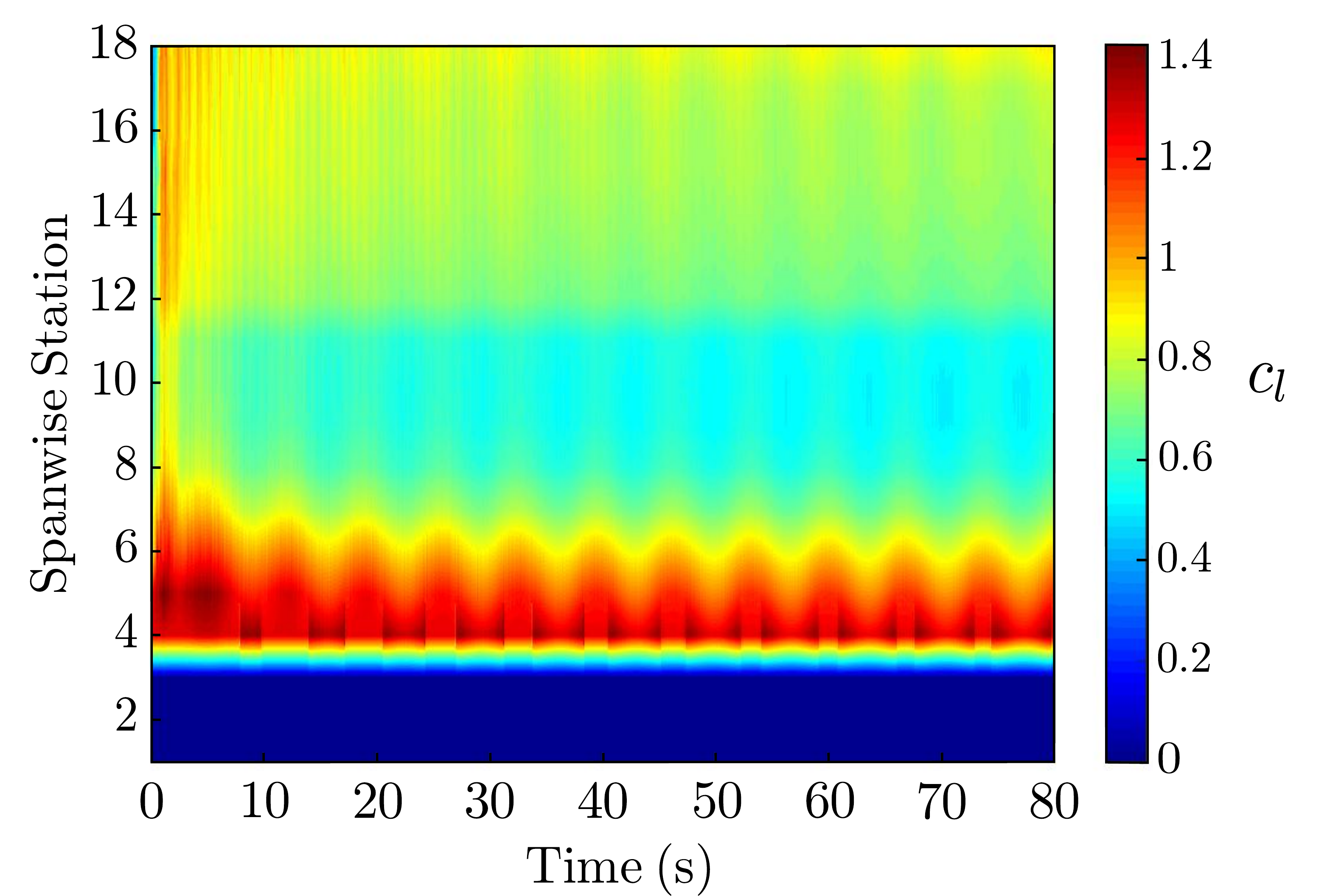}
            \caption[]%
            {Lift coefficient}    
            
        \end{subfigure}
        \hfill
        \begin{subfigure}[b]{0.475\textwidth}  
            \centering 
            \includegraphics[scale=0.225, trim=50 0 90 0]{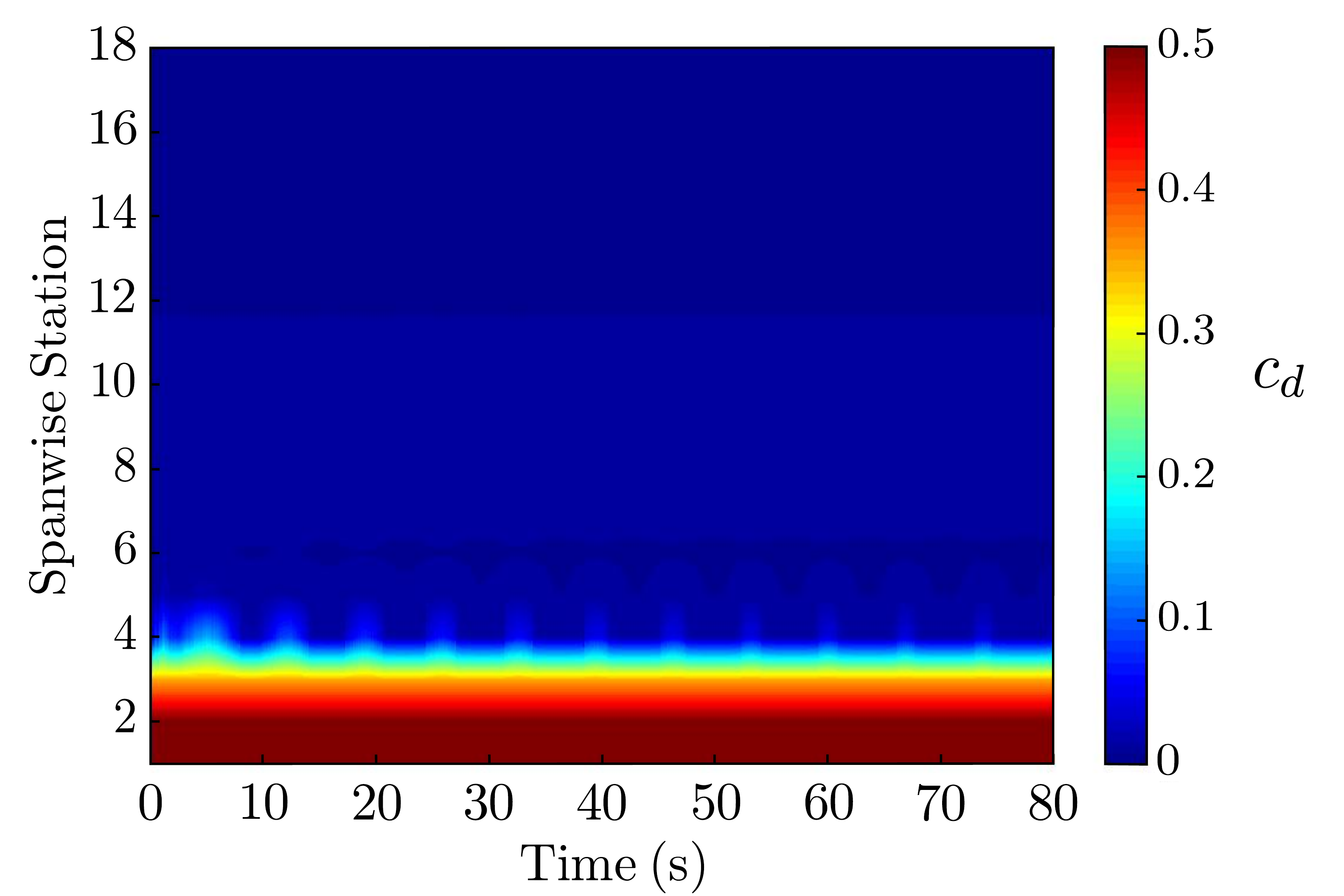}
            \caption[]%
            {Drag coefficient}    
            
        \end{subfigure}
        \vskip\baselineskip
        \begin{subfigure}[b]{0.475\textwidth}   
            \centering 
            \includegraphics[scale=0.225, trim=90 0 50 0]{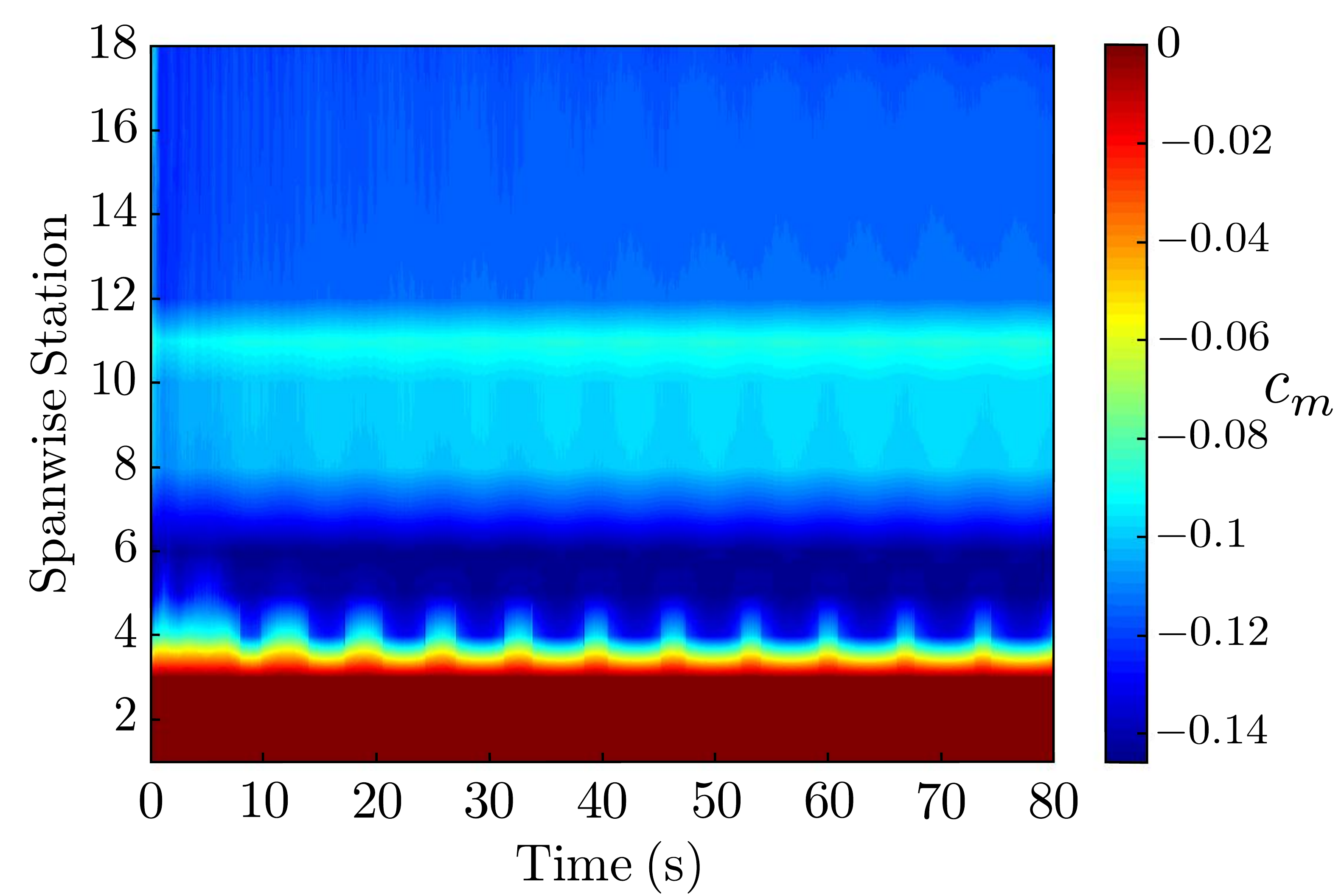}
            \caption[]%
            {Moment coefficient}    
            
        \end{subfigure}
        \caption[]
        {Time history of aerodynamic loads for blade 1 of the below-rated OC3 floating offshore wind turbine} 
        \label{BR_aerodynamics_fowt}
    \end{figure*}

\begin{figure*}[p!]
        \centering
        \begin{subfigure}[b]{0.475\textwidth}
            \centering
            \includegraphics[scale=0.45, trim=90 210 50 250]{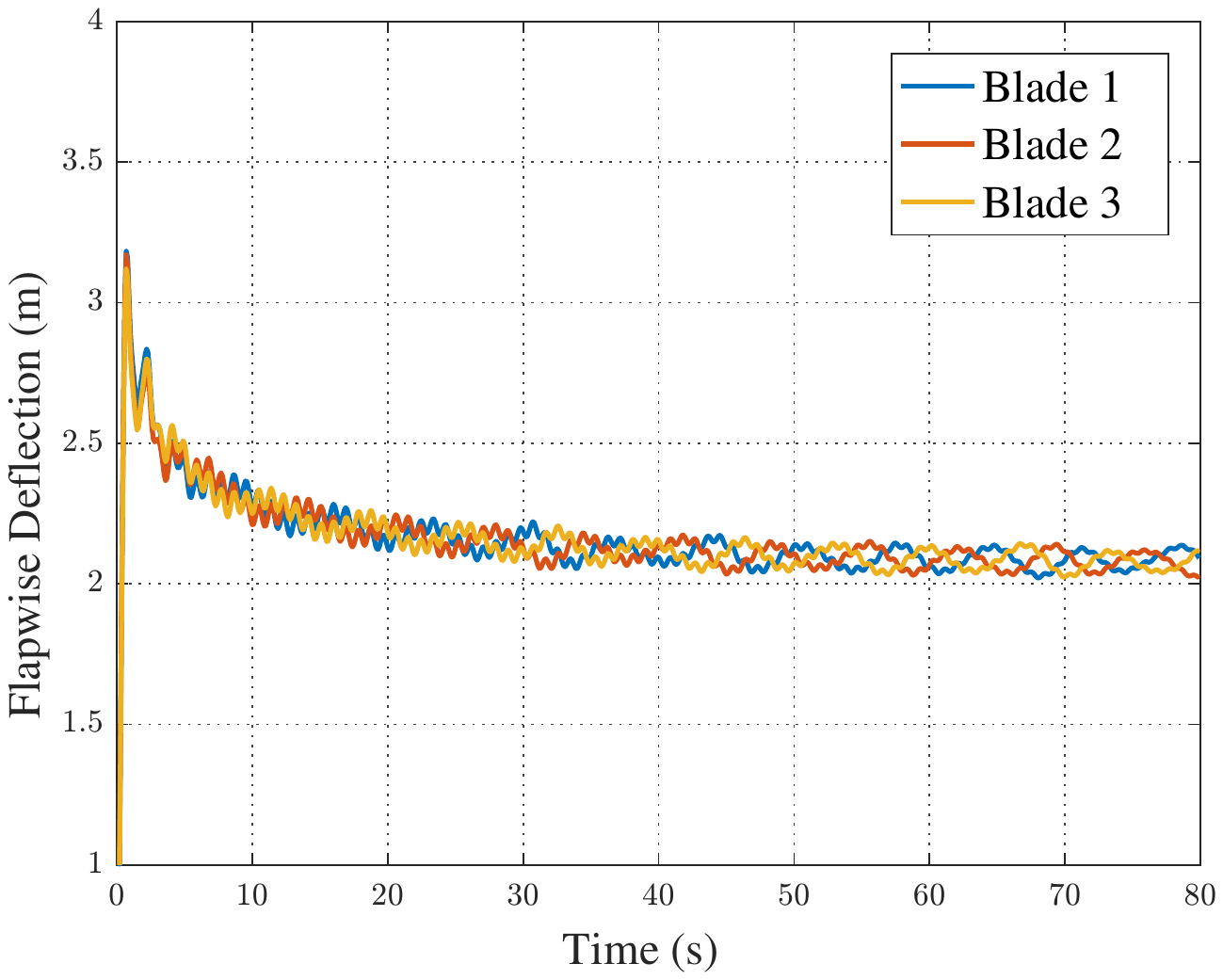}
            \caption[]%
            {{Flapwise deflection}}    
            \label{BR_flap_defl}
        \end{subfigure}
        \hfill
        \begin{subfigure}[b]{0.475\textwidth}  
            \centering 
            \includegraphics[scale=0.45, trim=90 210 25 250]{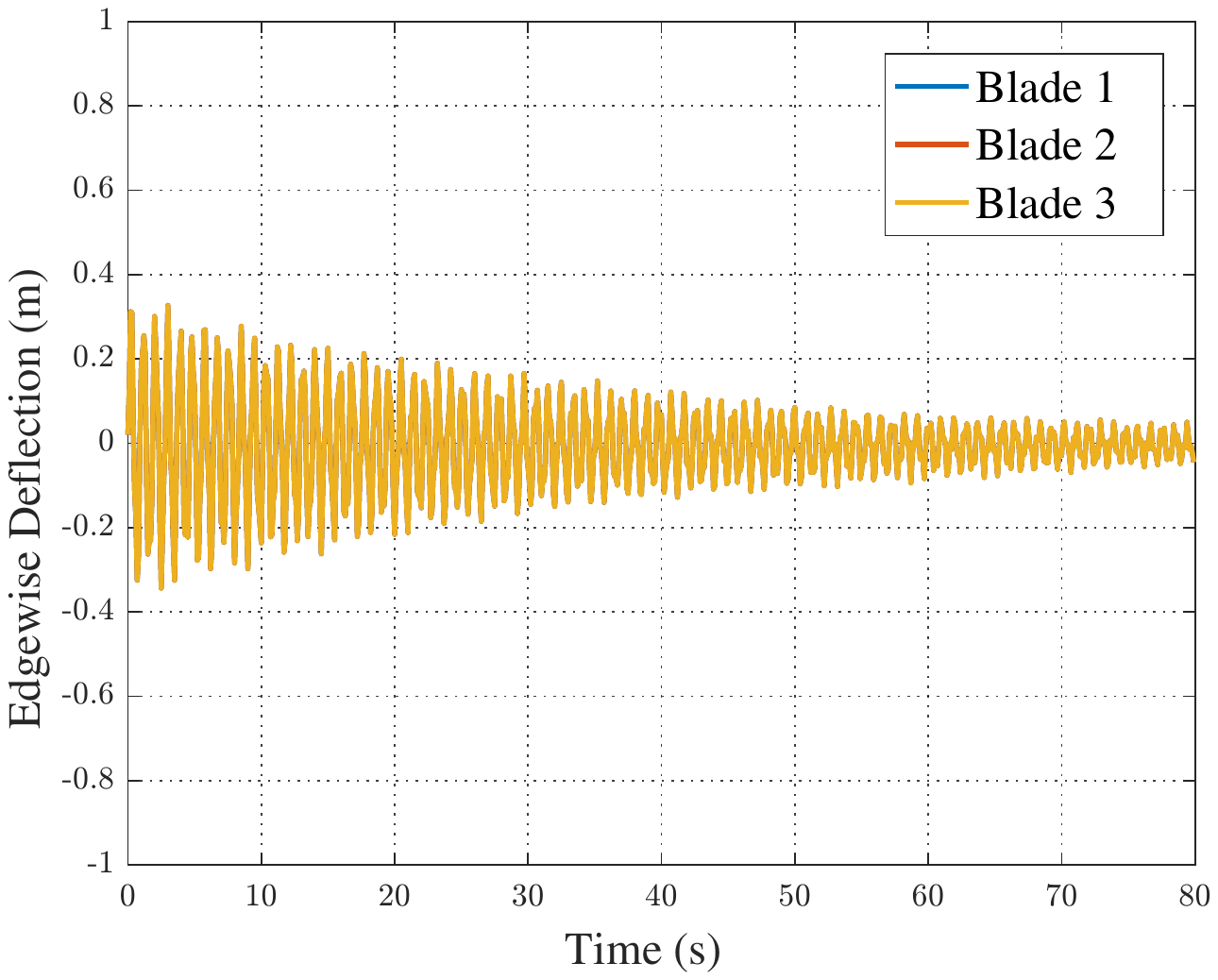}
            \caption[]%
            {{\small Edgewise deflection}}    
            \label{BR_edge_defl}
        \end{subfigure}
        \vskip\baselineskip
        \begin{subfigure}[b]{0.475\textwidth}   
            \centering 
            \includegraphics[scale=0.45, trim=90 210 50 220]{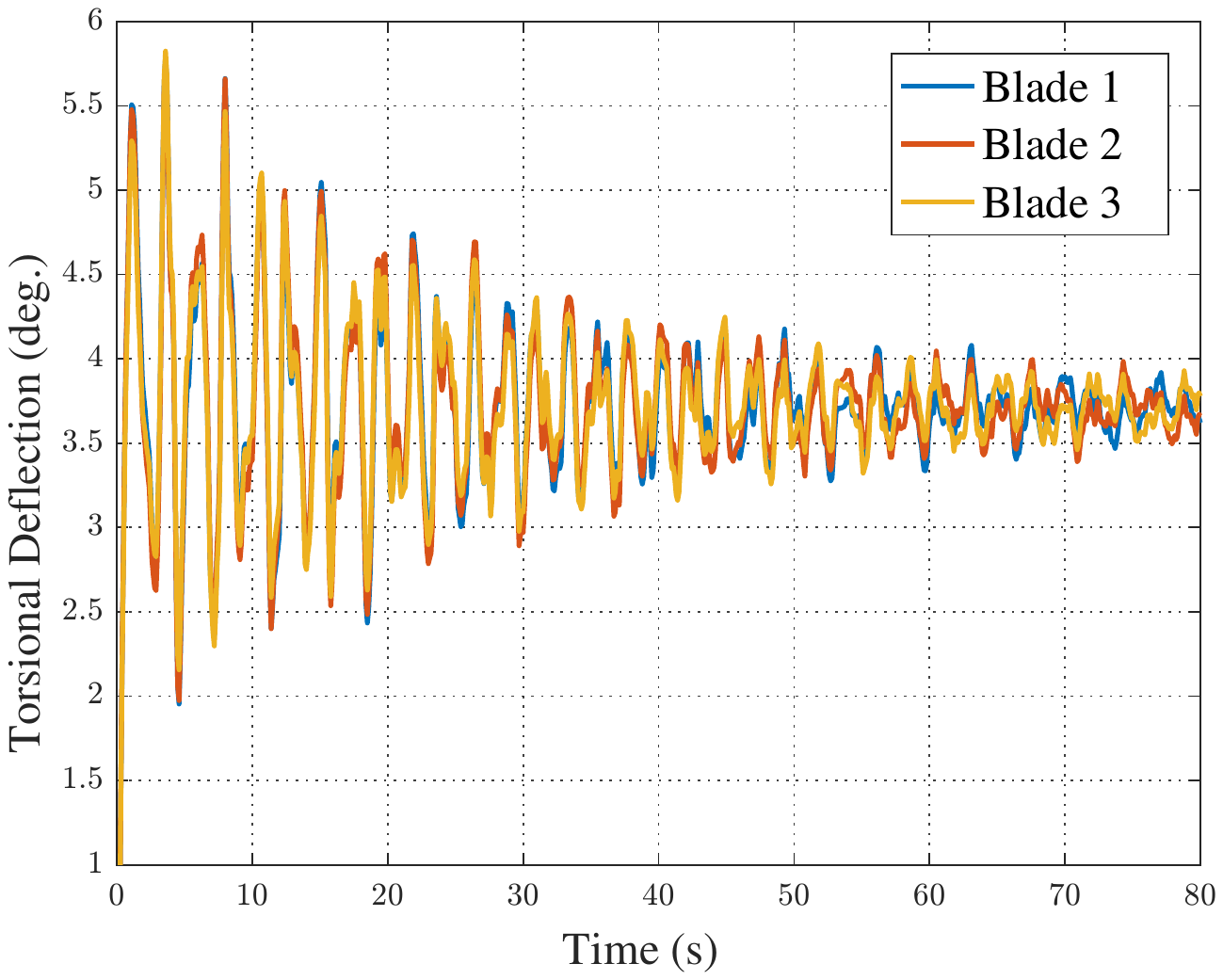}
            \caption[]%
            {{\small Torsional deflection}}    
            \label{BR_tor_defl}
        \end{subfigure}
        \caption[Time history of FOWT rotor-blade tip deflections at below-rated conditions]
        {Time history of rotor-blade tip deflections at below-rated operational conditions} 
        \label{BR_FOWT_blade_deflect}
    \end{figure*}

The aerodynamic loading for an individual blade shown in Fig.~\ref{BR_aerodynamics_fowt} also does not show any significant influence of the prescribed rigid-body motion, but a periodic loading is observed in the lift, drag, and moment coefficients. This periodic loading is due to the $5^{\circ}$ rotor plane tilt called for in the NREL 5-MW reference wind turbine configuration, which causes changes in the local blade velocities as each individual blade moves around the tilted plane. Since each blade is evenly spaced in the azimuth direction, the local change in velocity due to the rotor plane tilt generates out-of-phase loading in time. This out-of-phase periodic loading is highlighted in Fig.~\ref{BR_flap_defl}, which shows out-of-phase blade deflection as each blade passes through the tilted rotor-plane. The flapwise deflections approach a converged periodic state at around 60 s into the simulation, while both the edgewise and torsional deflections have not yet fully reached a steady or periodic state by 60 s into the simulation, as shown in Fig.~\ref{BR_edge_defl} and Fig.~\ref{BR_tor_defl}. The edgewise deflection experiences a very low-amplitude and high-frequency response compared to the flapwise deflection, which is expected because of its considerable stiffness, marginal loading in the edgewise direction, and the modeling assumptions made in \cite{rodriguez_JRE_p1} that did not account for modal coupling or lift contributions in the edgewise direction due to blade twist. Even though the transient frequency content of the edgewise deflection has not yet fully settled by the end of the simulation, it is clear to see in Fig.~\ref{BR_edge_defl} that the edgewise behavior approaches a zero mean deflection as its steady state solution. Similarly, the torsional deflection in Fig.~\ref{BR_tor_defl} does not finish converging by the end of the simulation, but the trend clearly approaches a periodic solution around $3.75^{\circ}$ in torsional deflection. 

\begin{figure*}[p!]
        \centering
        \begin{subfigure}[b]{0.475\textwidth}
            \centering
            \includegraphics[scale=0.45, trim=90 210 50 250]{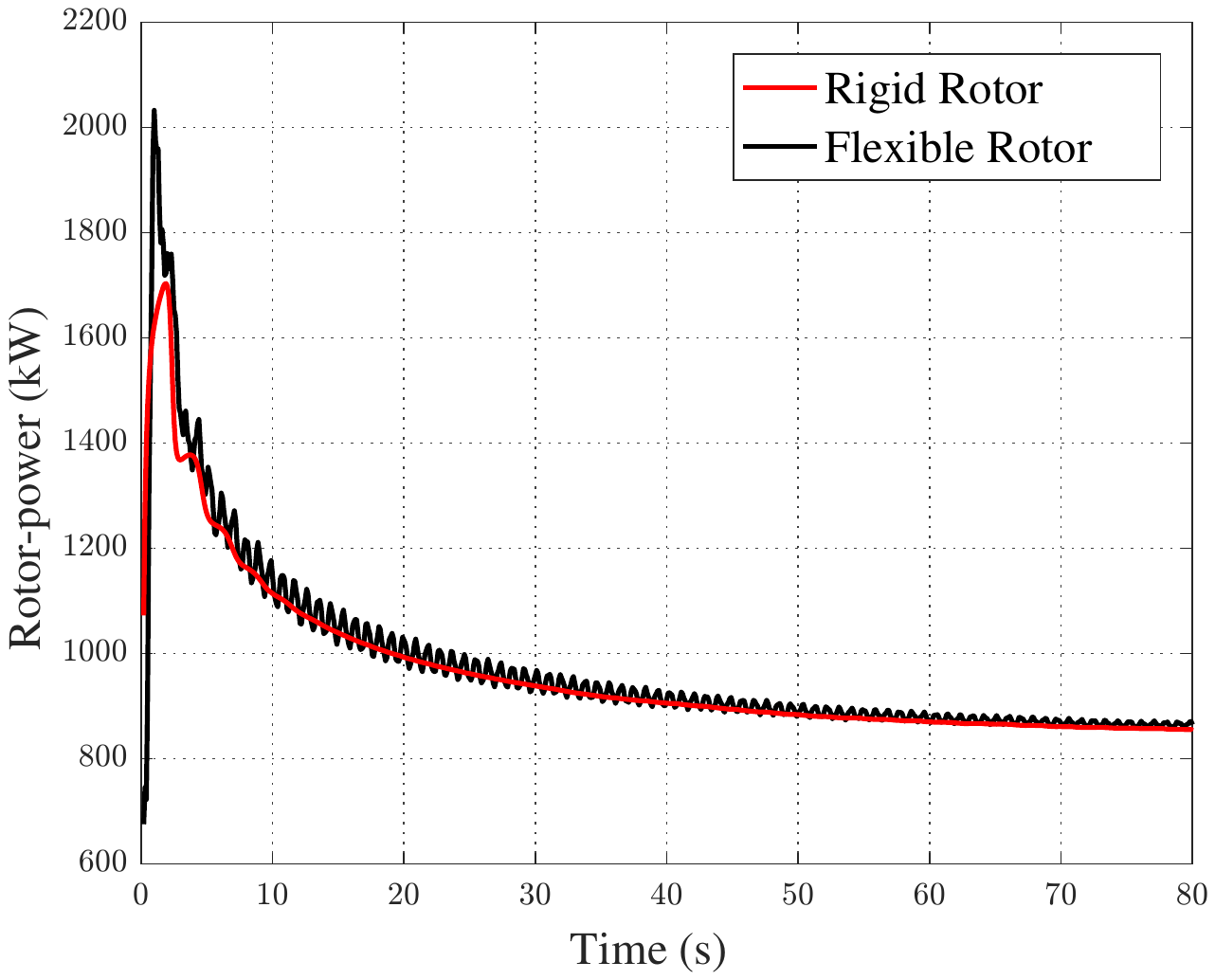}
            \caption[]%
            {{Rotor power}}    
            
        \end{subfigure}
        \hfill
        \begin{subfigure}[b]{0.475\textwidth}  
            \centering 
            \includegraphics[scale=0.45, trim=90 210 25 250]{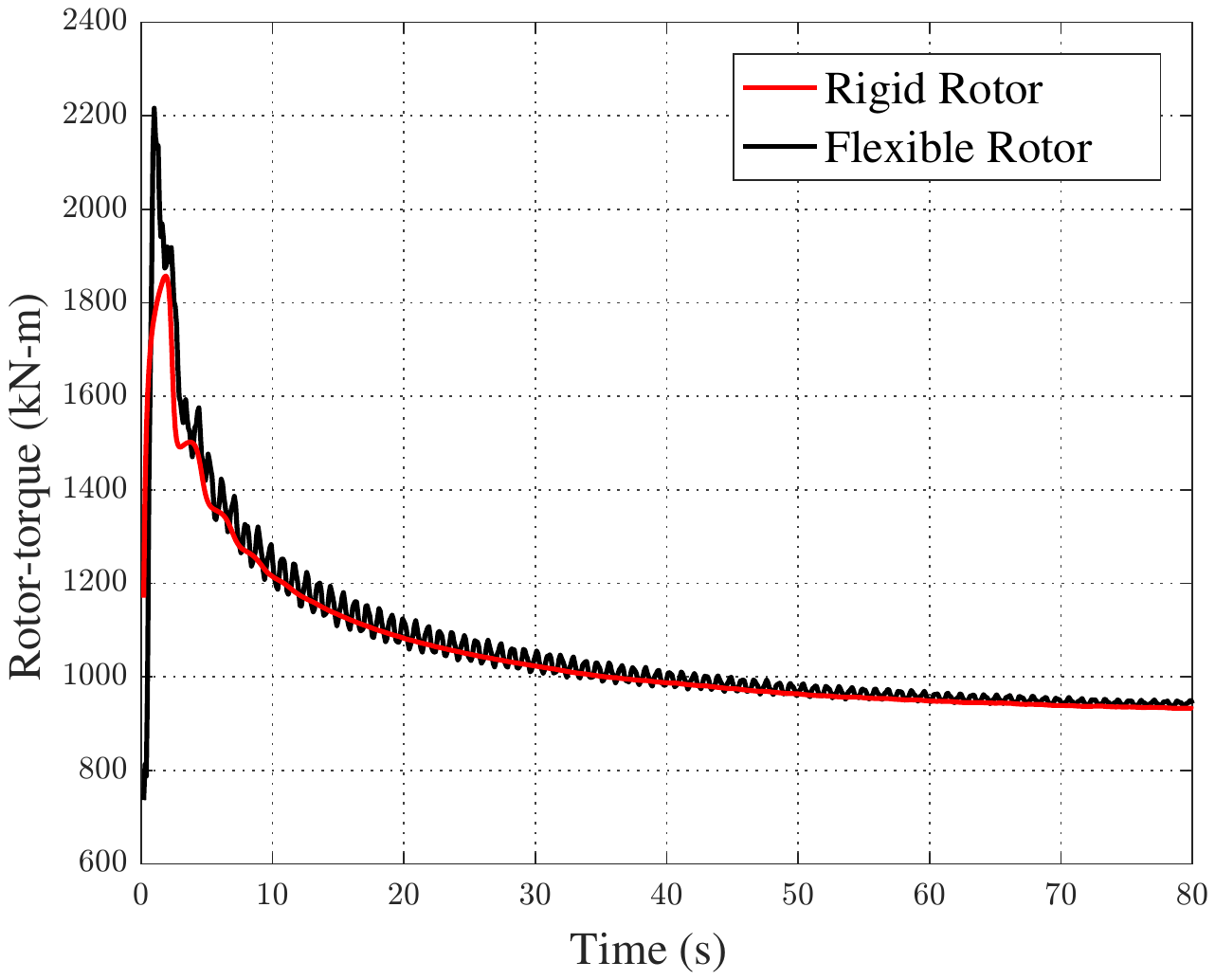}
            \caption[]%
            {{\small Rotor torque}}    
            
        \end{subfigure}
        \vskip\baselineskip
        \begin{subfigure}[b]{0.475\textwidth}   
            \centering 
            \includegraphics[scale=0.45, trim=90 210 50 220]{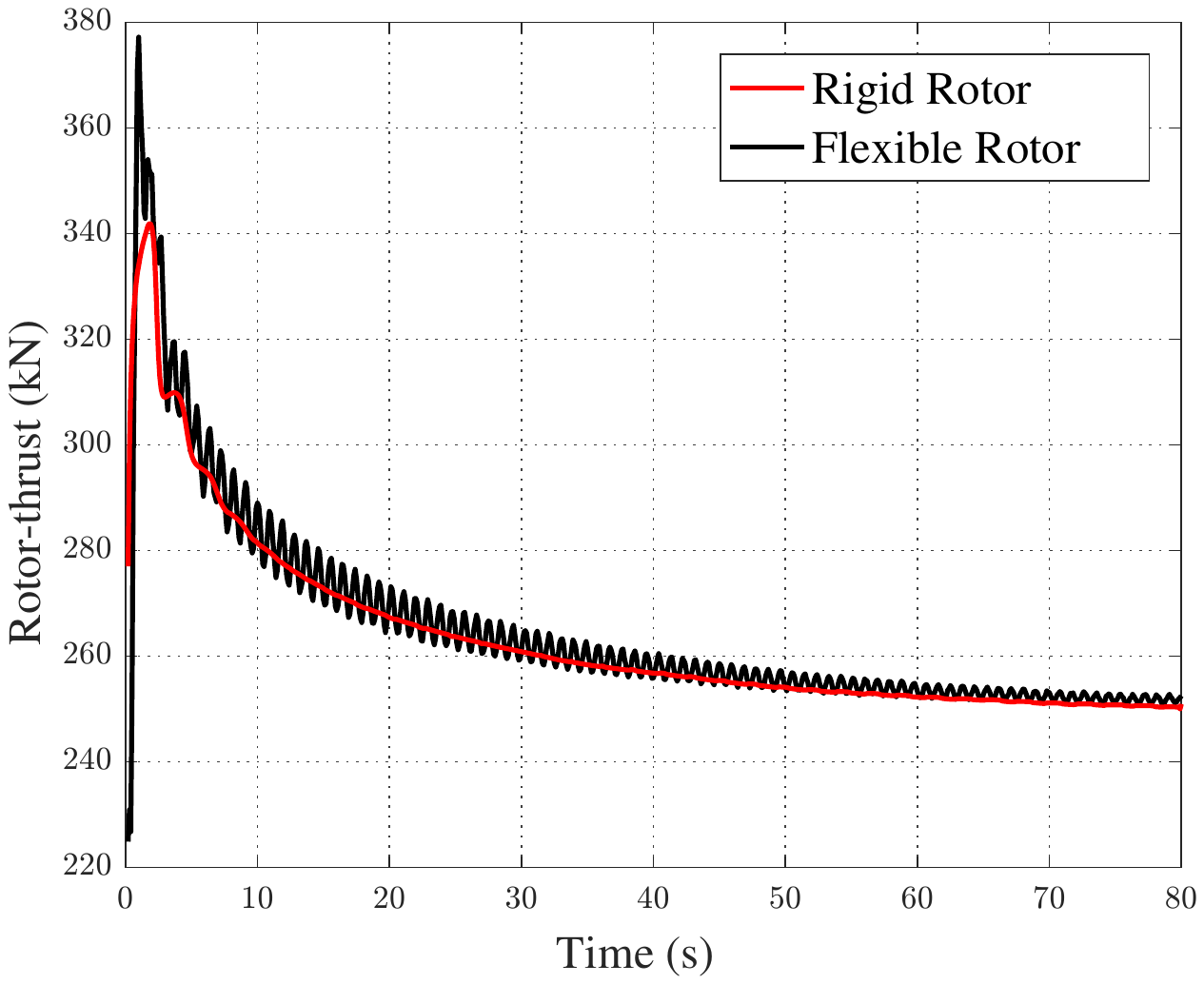}
            \caption[]%
            {{\small Rotor thrust}}    
            
        \end{subfigure}
        \caption[FOWT rotor performance metrics at below-rated operational conditions]
        {Rigid and flexible time history of rotor operational performance at below-rated operational conditions } 
        \label{BR_FOWT_perf_metrics}
    \end{figure*}

The flexible rotor performance of the below-rated FOWT case was compared to its rigid rotor counterpart. As shown in Fig.~\ref{BR_FOWT_perf_metrics}, the flexible rotor generates similar results as the rigid rotor with the exception of performance fluctuation. This performance fluctuation seems to be a byproduct of the blade deflection, specifically the flapwise deflection. The rigid rotor performance shares the same trends as the flexible rotor, such that its performance approximates the mean of the flexible rotor performance. Finally, there is no appreciable influence of the wave-induced motions from the offshore environment on the rotor performance in the below-rated case.

\subsection{Rated Operation}
\begin{figure}[h!]
\center
\includegraphics[scale=0.125, trim=0cm 75cm 30cm 10cm,clip]{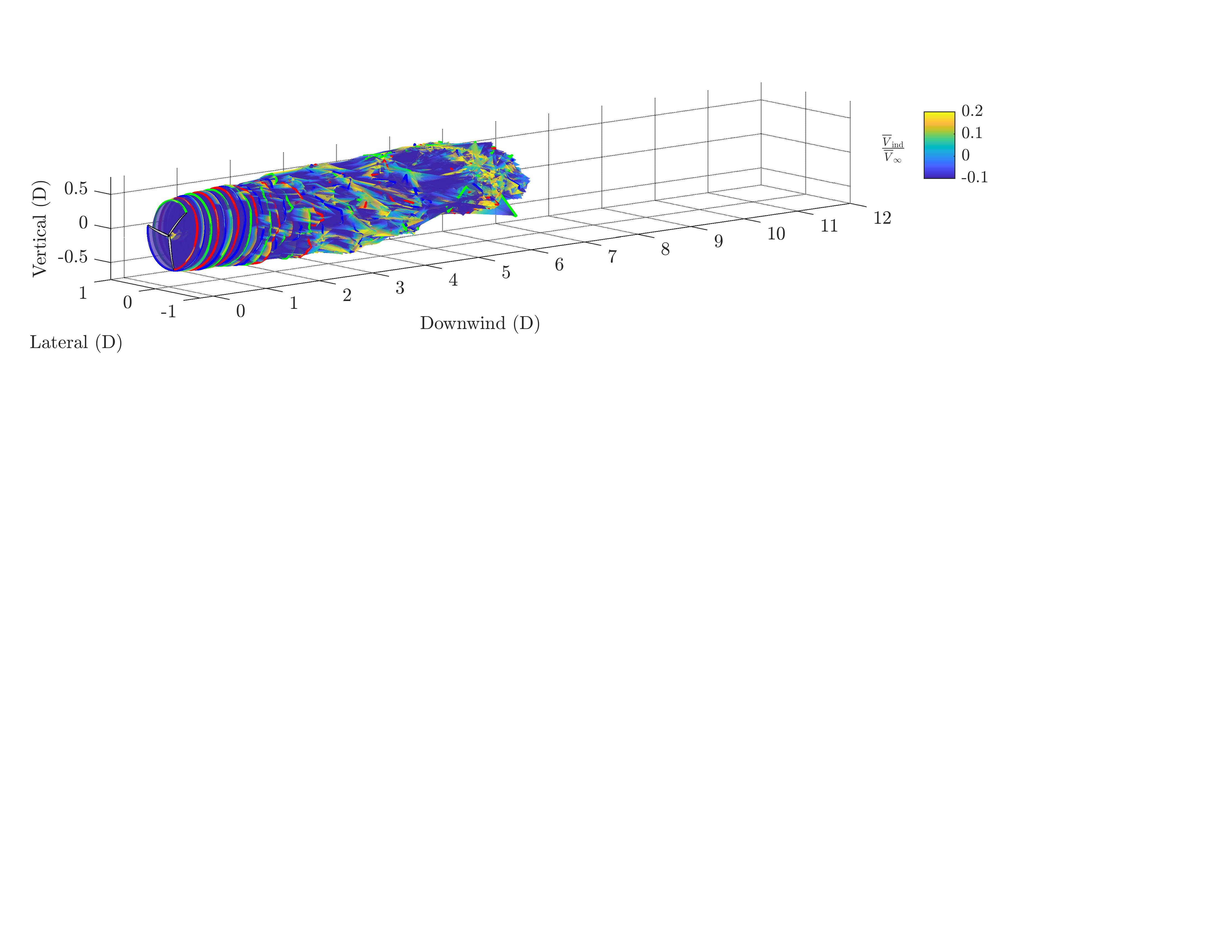}
\caption[FOWT wake shed from a flexible rotor at rated operational conditions]{Simulation of the wake shed from a flexible rotor at rated operational conditions. The magnitude of the induced velocity and the freestream velocity are $\bar{V}_{\rm ind}$ and $\bar{V}_{\infty}$, respectively. Spatial dimensions are scaled by the rotor diameter, $D$.}
\label{R_wake_flex_FOWT}
\end{figure}

The wake generated by the rated operational conditions of the floating offshore wind turbine is shown in Fig.~\ref{R_wake_flex_FOWT}. Coherence of the wake lasts about 1.5 diameters downstream before it begins to breakdown. The induced velocity also opposes strongly the inflow condition until about 1.5 diameters before the wake breaks down. The blade deflections shown in Fig.~\ref{R_wake_flex_FOWT} are also much more noticeable than the prior below-rated wake. Due to the higher inflow, the wake has also convected downstream much further than the below-rated case. The wake has a stretched appearance at locations 4 and 6 diameters downstream of the rotor, which was generated from the activation of the rigid-body motion.

\begin{figure*}[p!]
        \centering
        \begin{subfigure}[b]{0.475\textwidth}
            \centering
            \includegraphics[scale=0.225, trim=120 0 50 0]{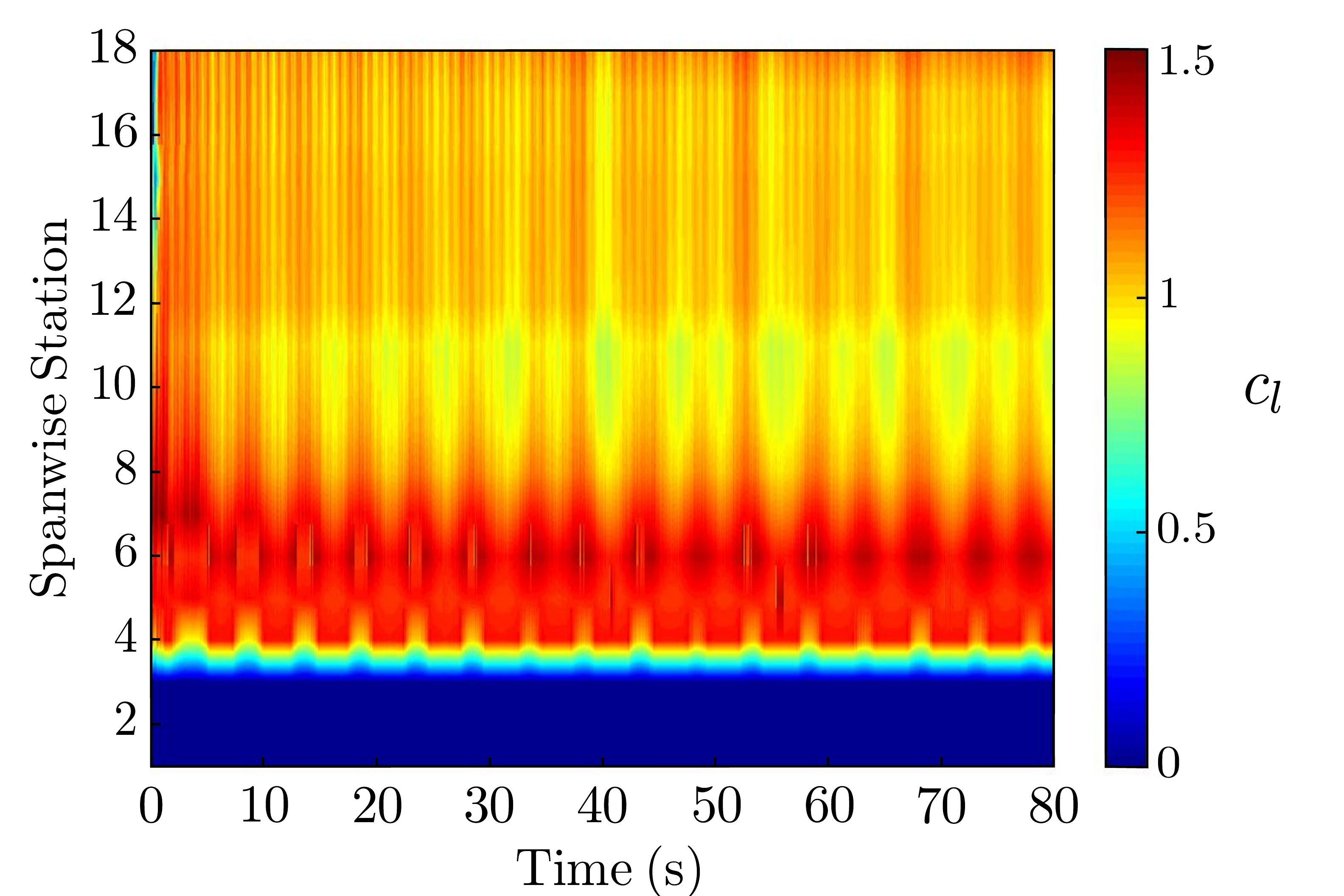}
            \caption[]%
            {Lift coefficient}    
            
        \end{subfigure}
        \hfill
        \begin{subfigure}[b]{0.475\textwidth}  
            \centering 
            \includegraphics[scale=0.225, trim=50 0 90 0]{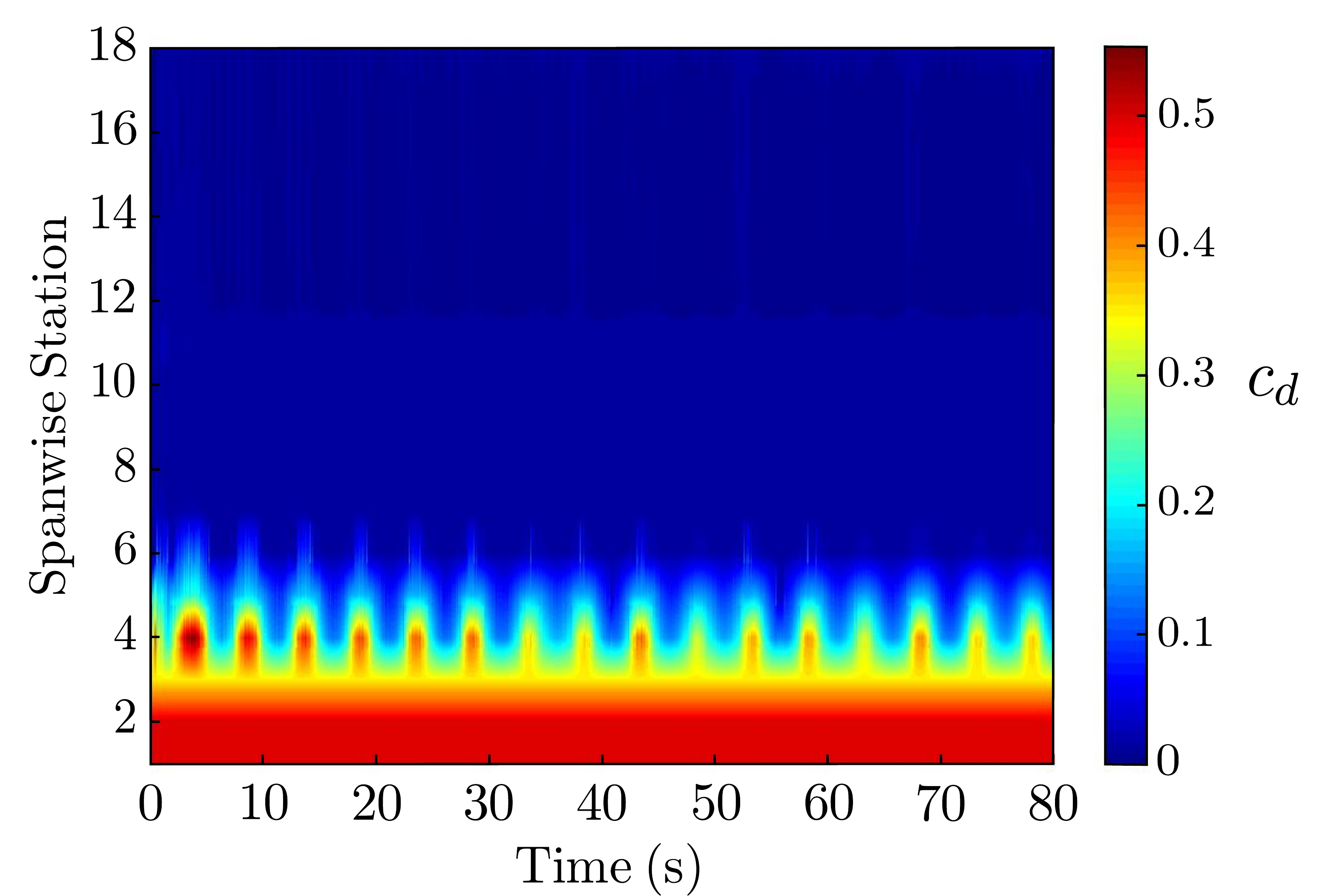}
            \caption[]%
            {Drag coefficient}    
            
        \end{subfigure}
        \vskip\baselineskip
        \begin{subfigure}[b]{0.475\textwidth}   
            \centering 
            \includegraphics[scale=0.225, trim=90 0 50 0]{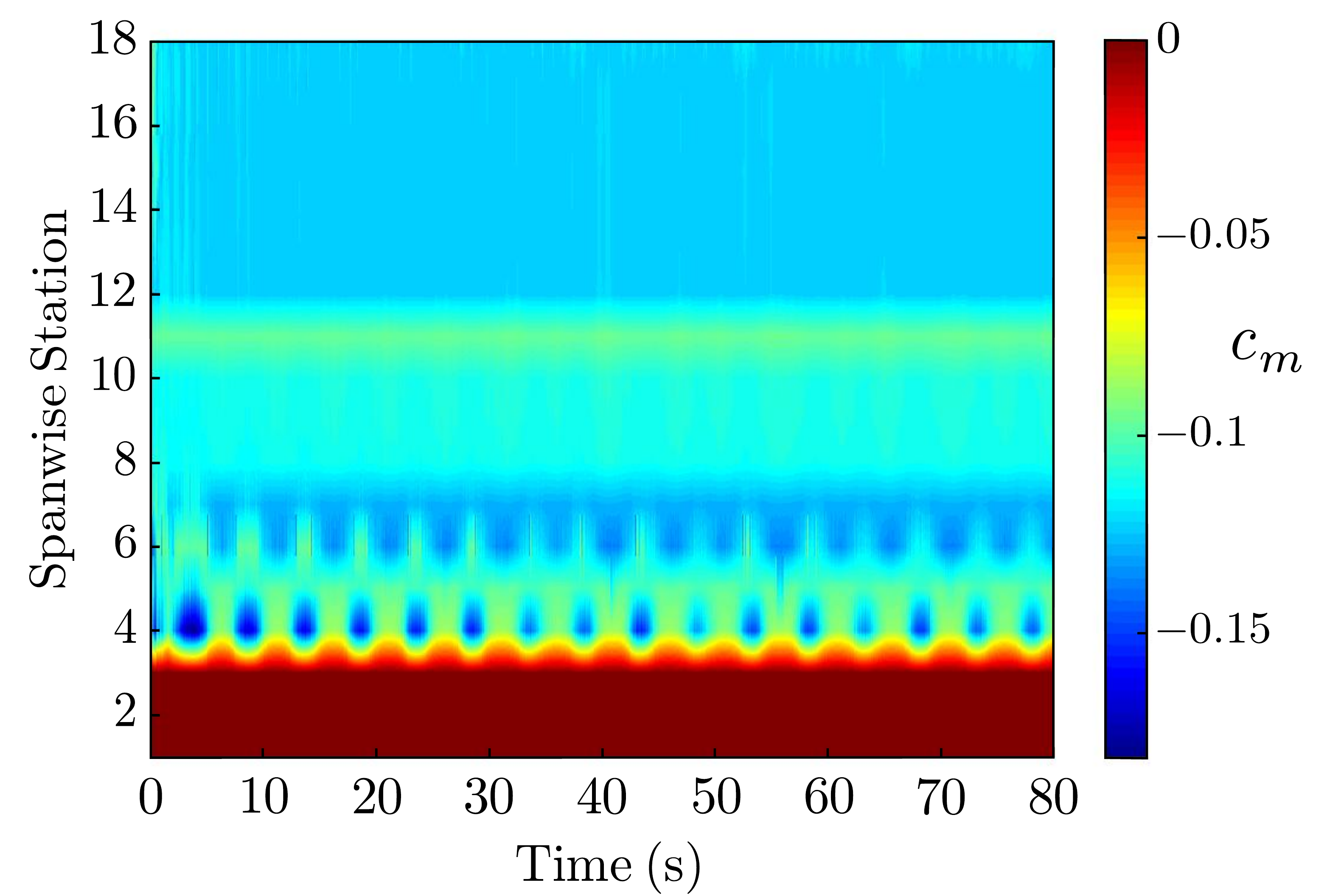}
            \caption[]%
            {Moment coefficient}    
            
        \end{subfigure}
        \caption[]
        {Time history of aerodynamic loads for blade 1 of the rated ITI floating offshore wind turbine} 
        \label{R_aerodynamics_fowt}
    \end{figure*}

\begin{figure*}[p!]
        \centering
        \begin{subfigure}[b]{0.475\textwidth}
            \centering
            \includegraphics[scale=0.45, trim=90 210 50 250]{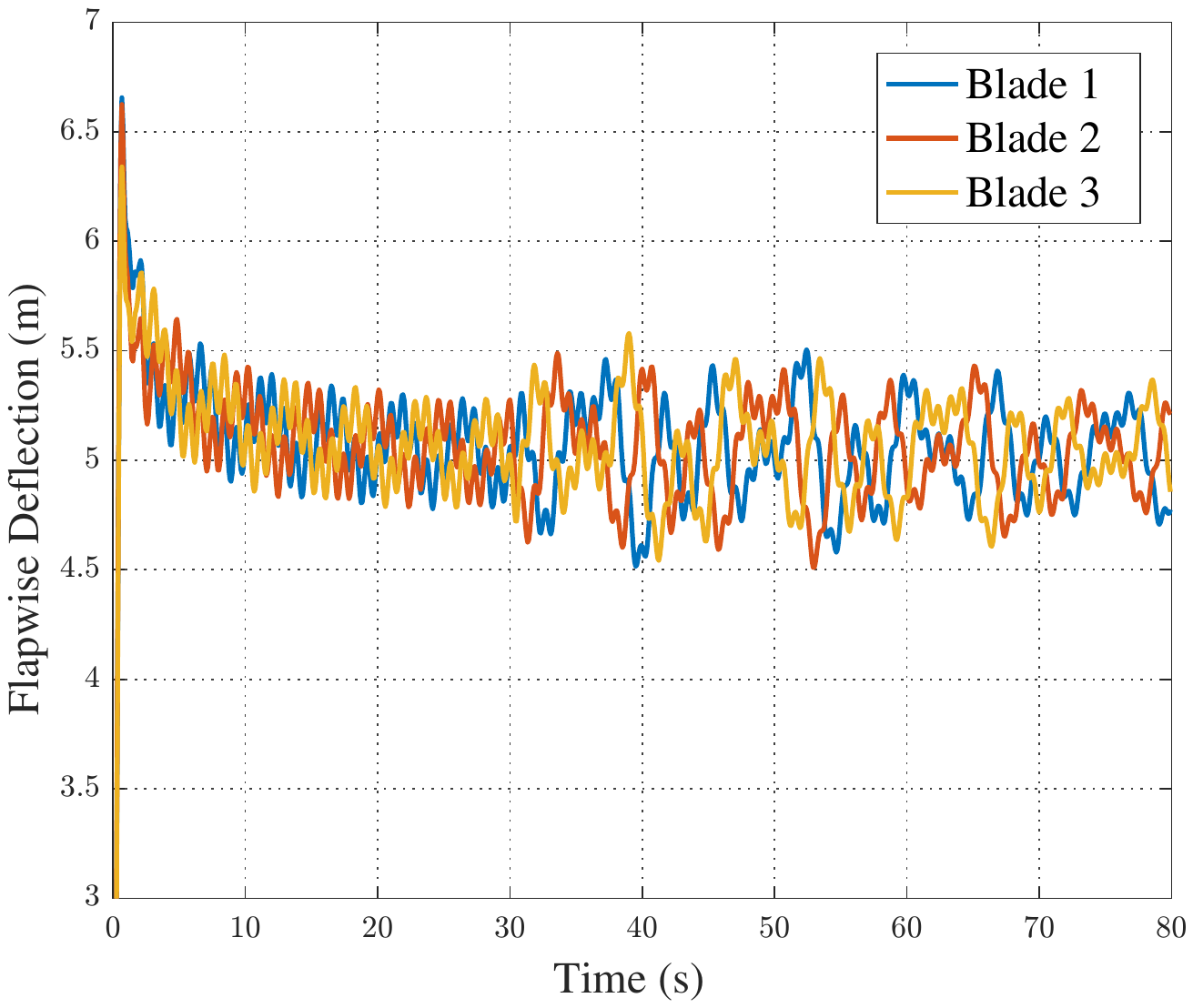}
            \caption[]%
            {{Flapwise deflection}}    
            \label{flap_r_fowt}
        \end{subfigure}
        \hfill
        \begin{subfigure}[b]{0.475\textwidth}  
            \centering 
            \includegraphics[scale=0.45, trim=90 210 25 250]{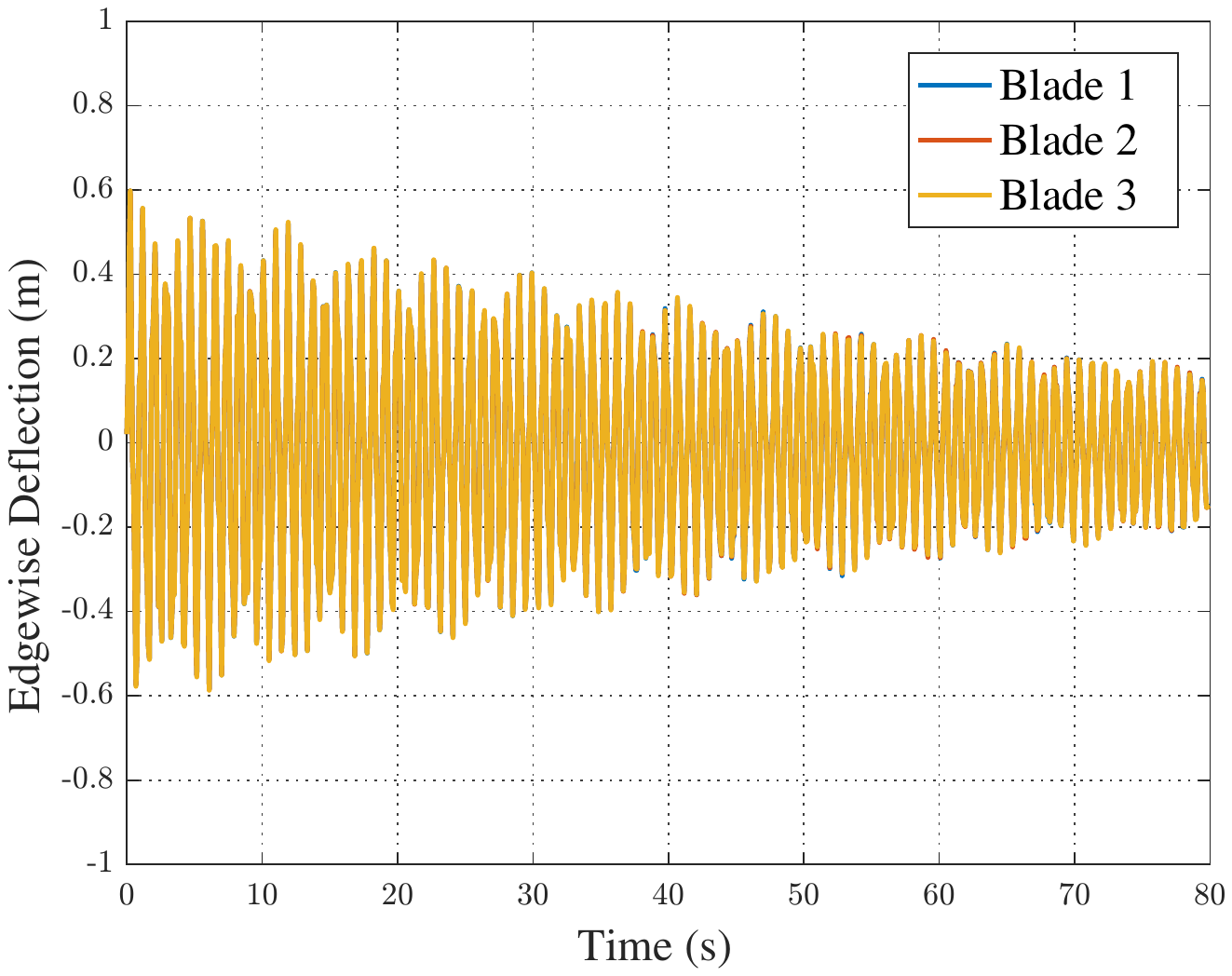}
            \caption[]%
            {{\small Edgewise deflection}}    
            \label{edge_r_fowt}
        \end{subfigure}
        \vskip\baselineskip
        \begin{subfigure}[b]{0.475\textwidth}   
            \centering 
            \includegraphics[scale=0.45, trim=90 210 50 220]{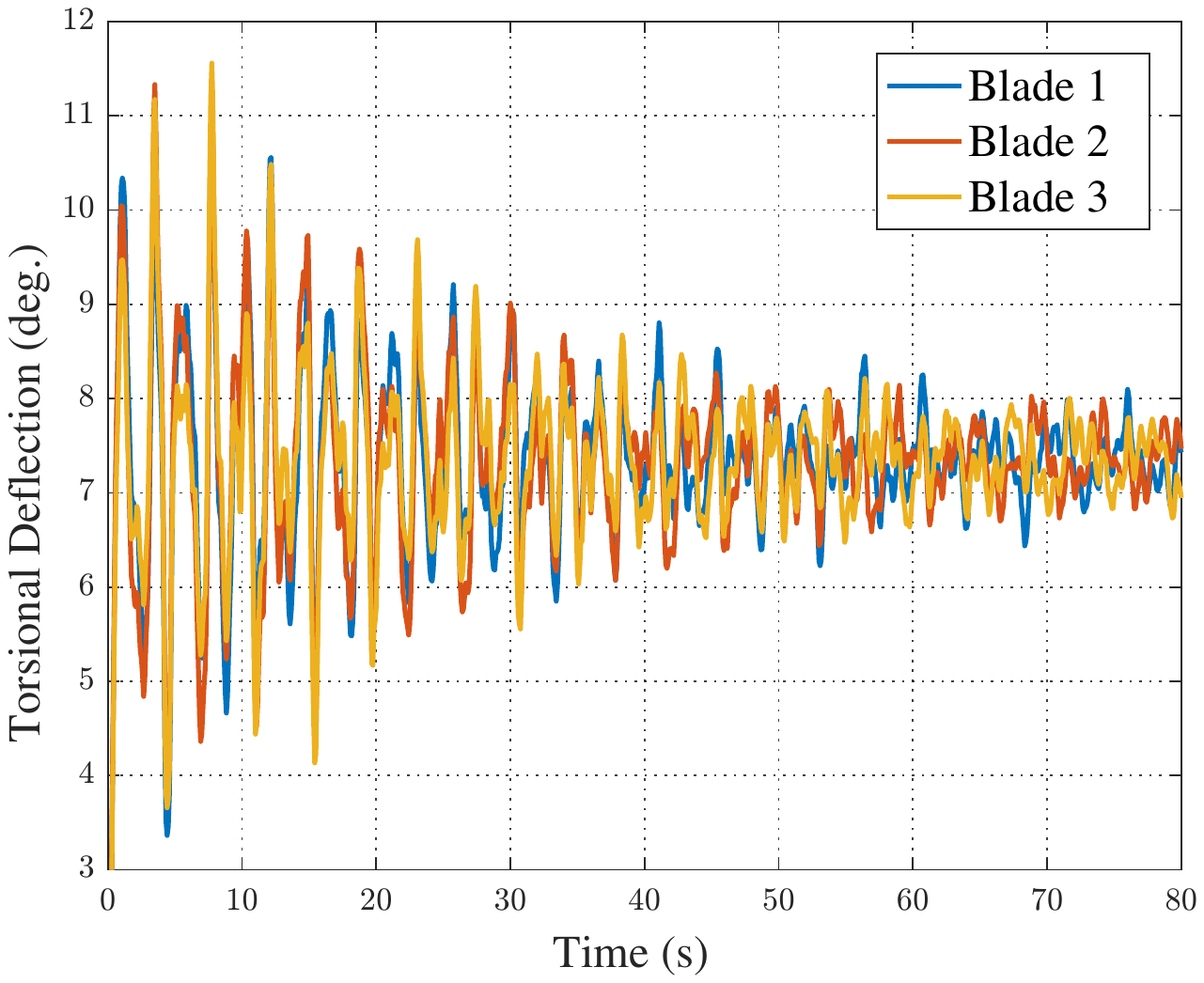}
            \caption[]%
            {{\small Torsional deflection}}    
            \label{tor_r_fowt}
        \end{subfigure}
        \caption[Time history of FOWT rotor-blade tip deflections at rated conditions]
        {Time history of rotor-blade tip deflections at rated operational conditions} 
        \label{R_blade_deflect_fowt}
    \end{figure*}

Unlike the below-rated operation, the aerodynamic loading in Fig.\ref{R_aerodynamics_fowt} for rated operation shows very notable impact from the wave-induced rotor motion. Specifically, the lift coefficient experiences high peaks across the blade span, and the drag and moment coefficients experience marginal drops at the corresponding rigid-body motion velocity peaks and troughs from Fig.~\ref{FOWT_platform_timehistories}. This periodic loading also directly impacts the blade deflection, as shown in Fig.~\ref{R_blade_deflect_fowt}. The flapwise deflection is the most impacted due to the modeling assumption presented \cite{rodriguez_JRE_p1}, such that rigid-body pitching motions are modeled as inertial loads in the flapwise direction only. Figure \ref{flap_r_fowt} clearly shows immediate reaction to the activation of the rigid-body motions. The reaction of the blade behavior goes like the acceleration of the rigid-body motion superposed with the blade-passing frequency occurring from the rotor-plane tilt. The edgewise deflection does not demonstrate the same behavior that was shown by the flapwise deflection because of the marginal loading, considerable stiffness of the blade in the edgewise direction, and the modeling assumptions presented in \cite{rodriguez_JRE_p1} do not account for mode coupling or lift contributions in the edgewise direction due to blade twist. Furthermore, the equation of motions derived assume negligible impact from the rigid-body motions in the edgewise direction, so it is expected that wave-induced motions did not impact the edgewise deflections. Similarly, the moment coefficient is marginally impacted by the rigid-body motions, as the rotor blade equations of motions assume that the rigid-body motions would not impact that torsional degree-of-freedom. Thus, there is no notable impact on torsional deflection caused by wave-induced rotor motions. 

\begin{figure*}[p!]
        \centering
        \begin{subfigure}[b]{0.475\textwidth}
            \centering
            \includegraphics[scale=0.45, trim=90 210 50 250]{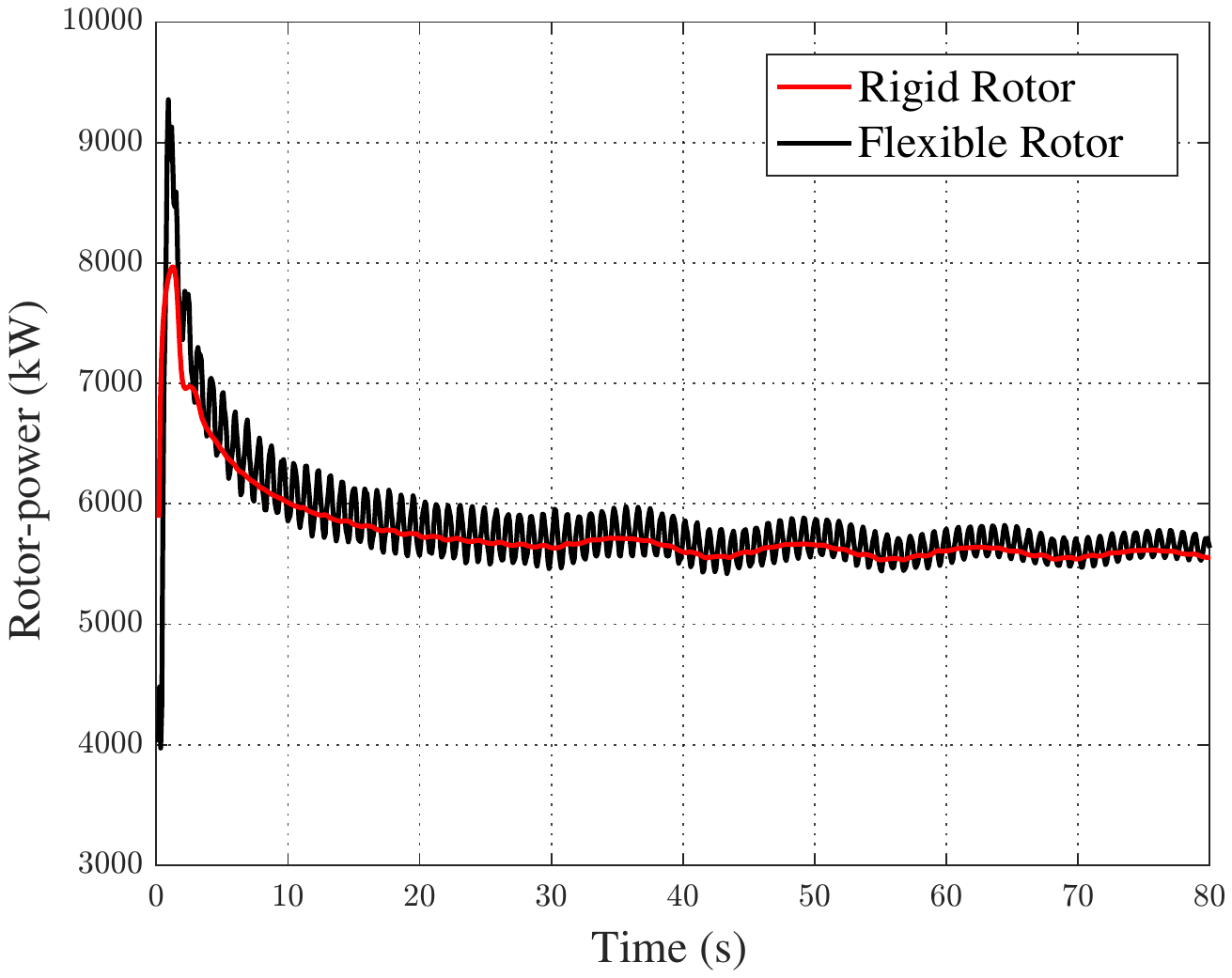}
            \caption[]%
            {{Rotor power}}    
            
        \end{subfigure}
        \hfill
        \begin{subfigure}[b]{0.475\textwidth}  
            \centering 
            \includegraphics[scale=0.45, trim=90 210 25 250]{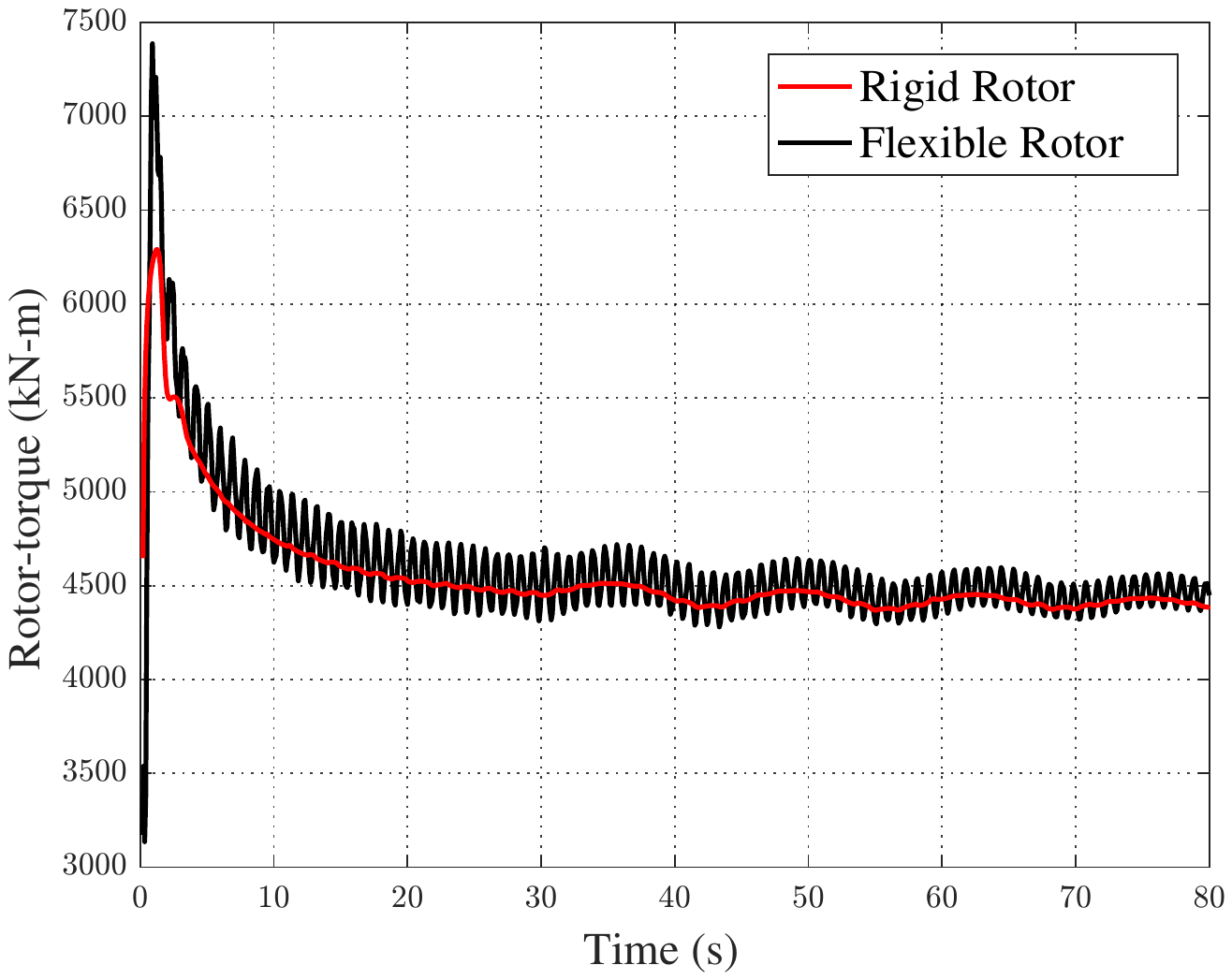}
            \caption[]%
            {{\small Rotor torque}}    
            
        \end{subfigure}
        \vskip\baselineskip
        \begin{subfigure}[b]{0.475\textwidth}   
            \centering 
            \includegraphics[scale=0.45, trim=90 210 50 220]{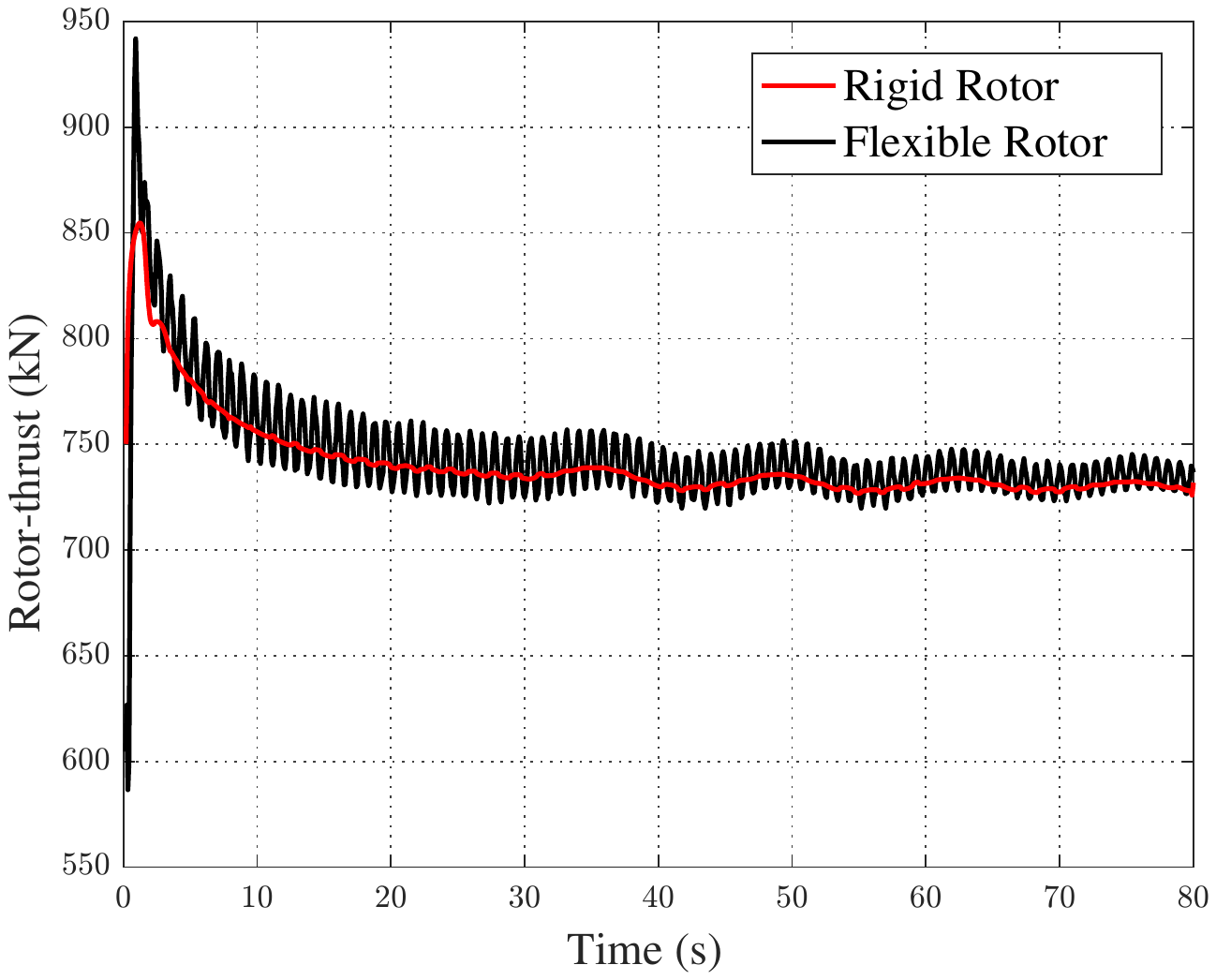}
            \caption[]%
            {{\small Rotor thrust}}    
            
        \end{subfigure}
        \caption[FOWT rotor performance metrics at rated operational conditions]
        {Rigid and flexible time history of rotor operational performance at rated operational conditions } 
        \label{R_perf_metrics_fowt}
    \end{figure*}

The flexible rotor performance of the rated FOWT case demonstrates impact from the wave induced rigid-body motions of the rotor, shown in Fig.~\ref{R_perf_metrics_fowt}. The power, torque, and thrust all have peaks and troughs at inflection points of the rigid-body motion velocities (i.e., the zero velocity). The performance metrics also fluctuate as a result to the flapwise blade deflection, as was discussed in the below-rated case.

\subsection{Above-rated Operation}
\begin{figure}[h!]
\center
\includegraphics[scale=0.125, trim=0cm 75cm 30cm 10cm,clip]{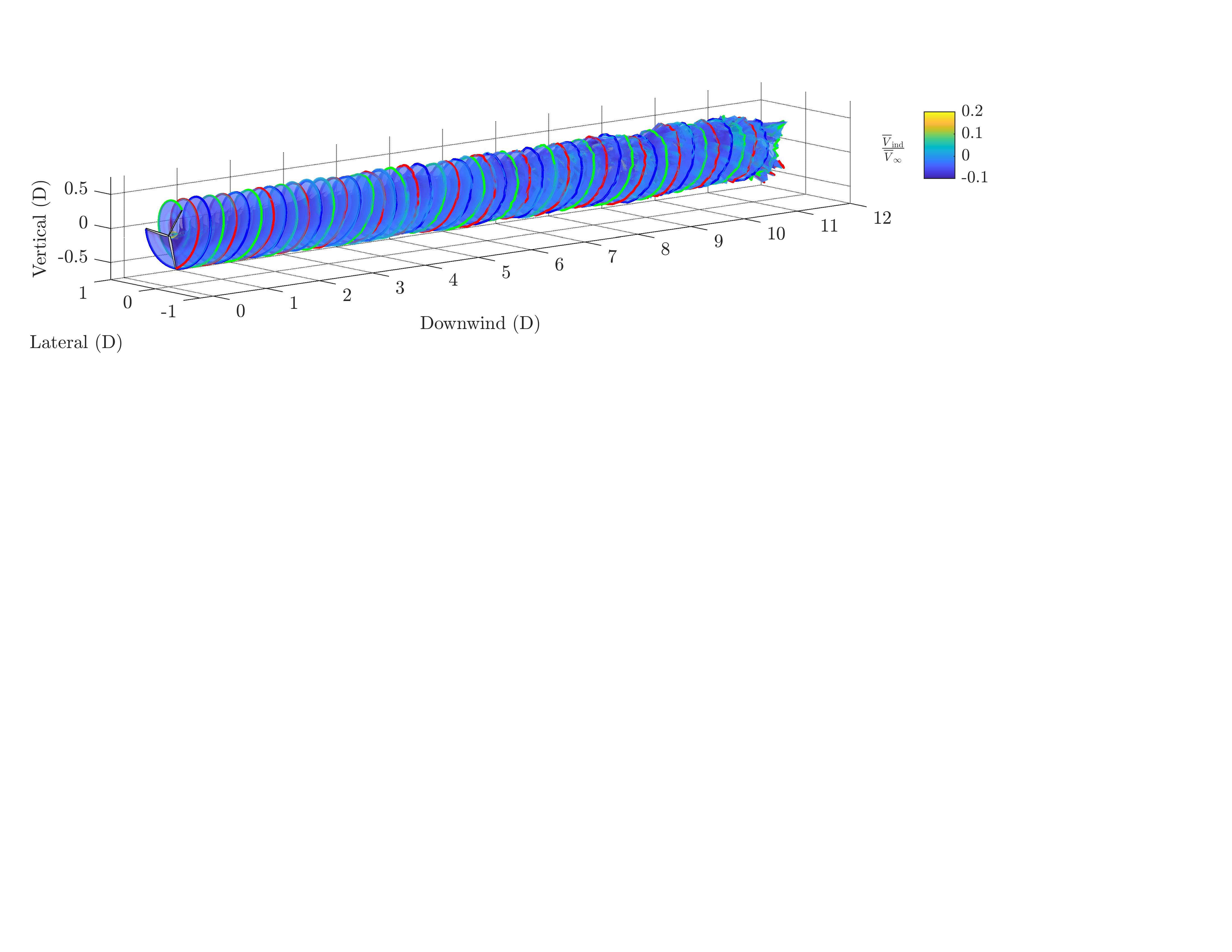}
\caption[FOWT wake simulation shed from a flexible rotor at above-rated operational conditions]{Simulation of the wake shed from a flexible rotor at above-rated operational conditions. The magnitude of the induced velocity and the freestream velocity are $\bar{V}_{\rm ind}$ and $\bar{V}_{\infty}$, respectively. Spatial dimensions are scaled by the rotor diameter, $D$.}
\label{AR_wake_flex_fowt}
\end{figure}

Finally, the wake generated by the above-rated inflow conditions is shown in Fig.~\ref{AR_wake_flex_fowt}. The above-rated conditions generate a rotor unlike the below-rated and rated wake due to its low tip-speed ratio and blade pitch that generate a relatively more stable wake than the former wakes, as is clearly seen in Fig.~\ref{AR_wake_flex_fowt}. These above-rated operational conditions thus make it difficult to note where the initial wake breakdown occurs. Furthermore, unlike the induced velocity field of the below-rated and rated case, the above-rated wake is not as strongly opposed to the inflow velocity. The rigid-body motions at the above-rated conditions have also impacted the generation of the wake: the helical pitch between adjacent tip-vortices and trailing vortex sheets vary as the wake convects downstream, i.e., trailing vortices and trailing vortex sheets are compressed and stretched according to the wave-induced motion.

\begin{figure*}[p!]
        \centering
        \begin{subfigure}[b]{0.475\textwidth}
            \centering
            \includegraphics[scale=0.225, trim=120 0 50 0]{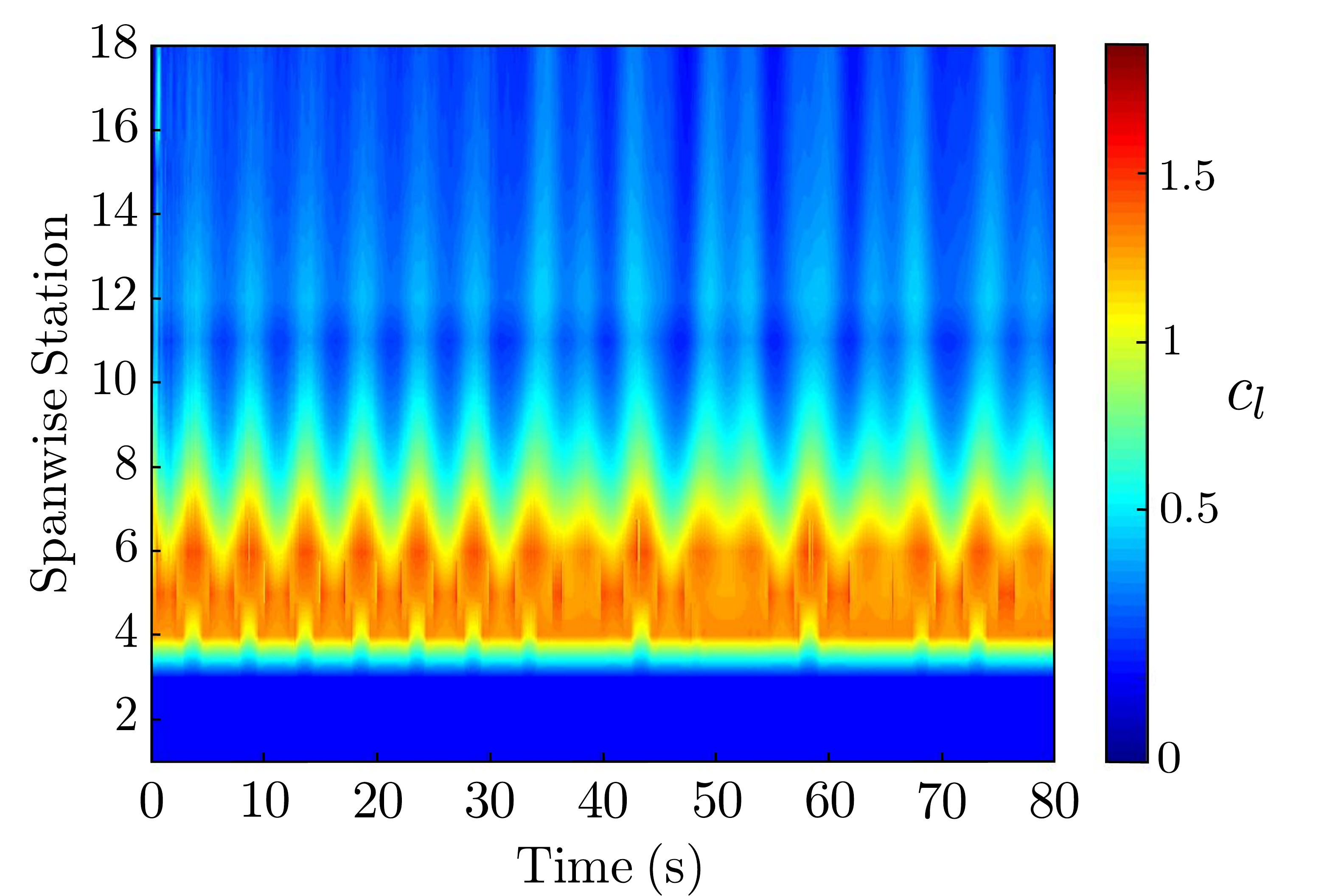}
            \caption[]%
            {Lift coefficient}    
            
        \end{subfigure}
        \hfill
        \begin{subfigure}[b]{0.475\textwidth}  
            \centering 
            \includegraphics[scale=0.225, trim=50 0 90 0]{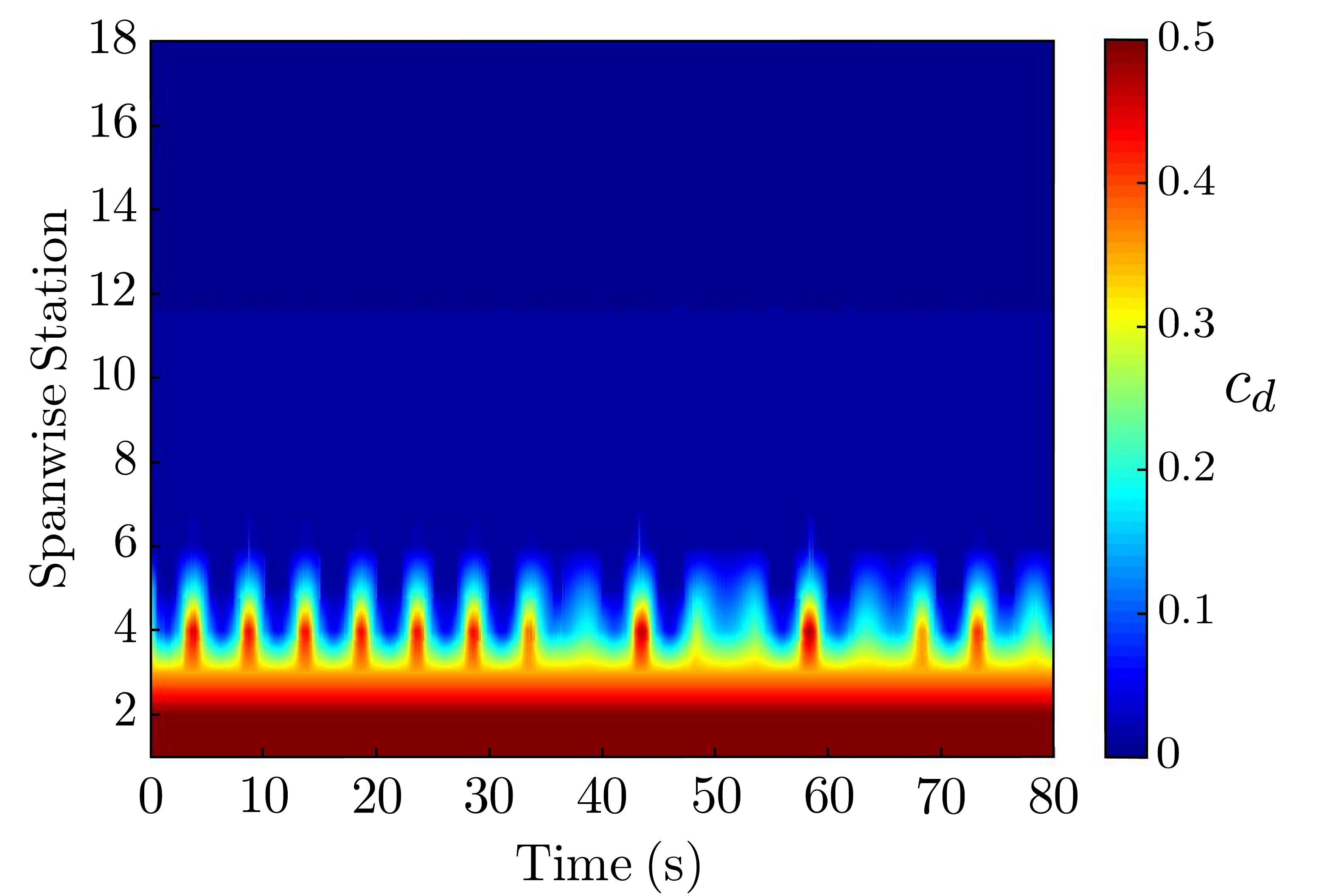}
            \caption[]%
            {Drag coefficient}    
            
        \end{subfigure}
        \vskip\baselineskip
        \begin{subfigure}[b]{0.475\textwidth}   
            \centering 
            \includegraphics[scale=0.225, trim=90 0 50 0]{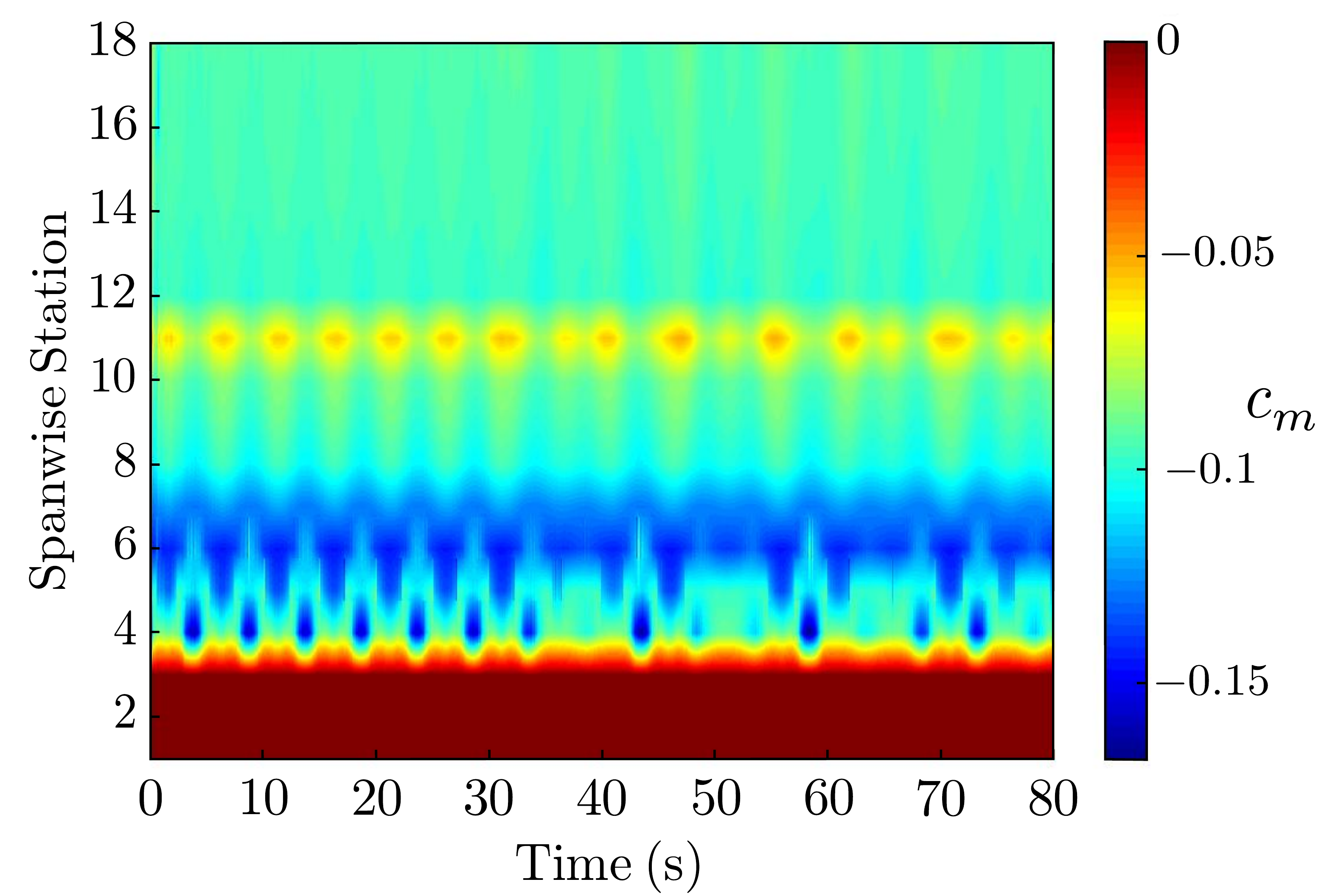}
            \caption[]%
            {Moment coefficient}    
            
        \end{subfigure}
        \caption[]
        {Time history of aerodynamic loads for blade 1 of the above-rated ITI floating offshore wind turbine} 
        \label{AR_aerodynamics_fowt}
    \end{figure*}

The above-rated condition highlights the rigid-body motions in aerodynamic loads the most out of the below-rated and rated operation. Figure \ref{AR_aerodynamics_fowt} clearly demonstrates the impact rigid-body motions have on aerodynamic loads. As seen in the rated operational state, periodic changes in the above-rated lift coefficients, drag, and moment coefficients correspond to the peaks and troughs of the rigid-body motion velocity presented in Fig.~\ref{AR_rbms}.

\begin{figure*}[p!]
        \centering
        \begin{subfigure}[b]{0.475\textwidth}
            \centering
            \includegraphics[scale=0.45, trim=90 210 50 250]{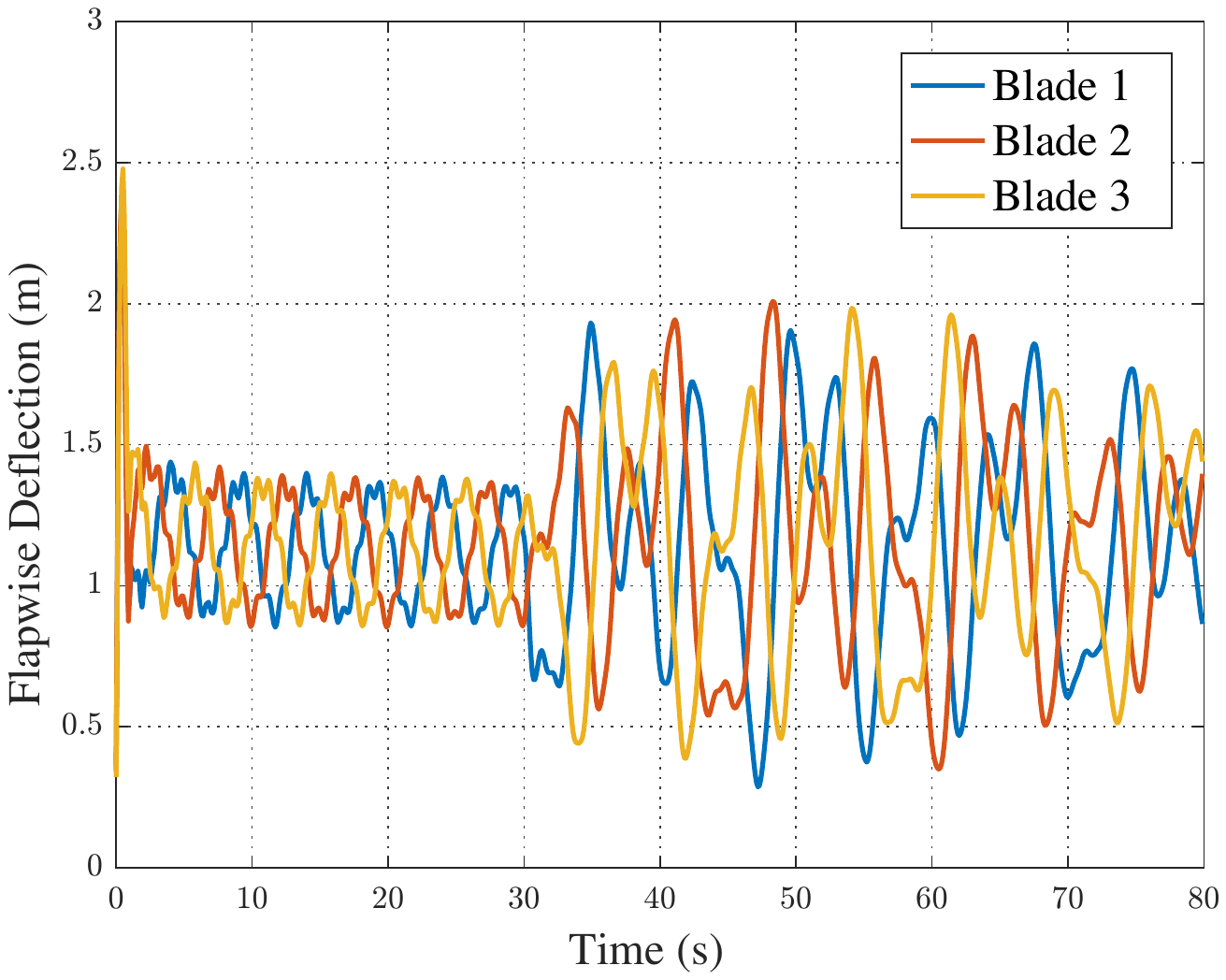}
            \caption[]%
            {{Flapwise deflection}}    
            
        \end{subfigure}
        \hfill
        \begin{subfigure}[b]{0.475\textwidth}  
            \centering 
            \includegraphics[scale=0.45, trim=90 210 25 250]{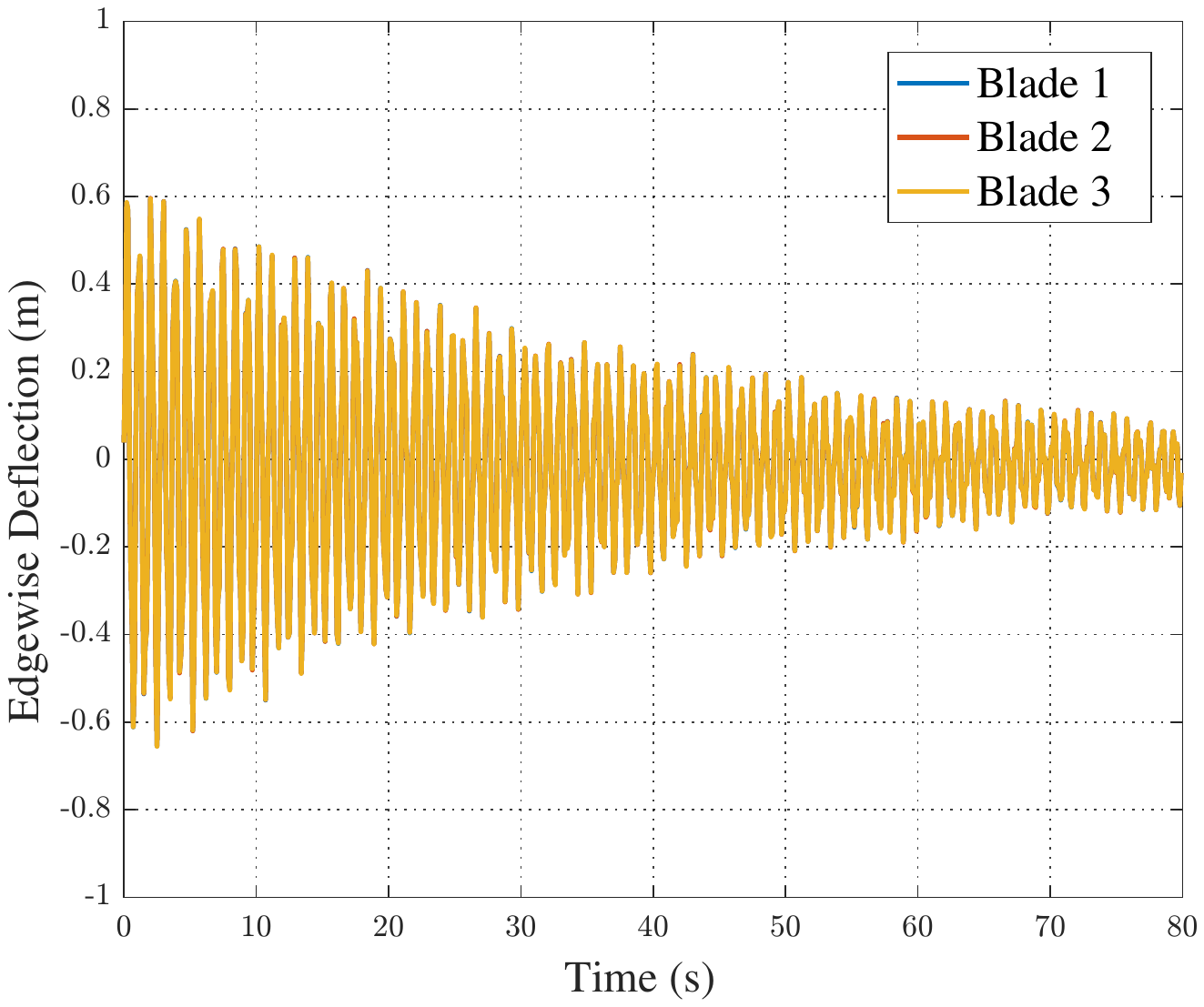}
            \caption[]%
            {{\small Edgewise deflection}}    
            
        \end{subfigure}
        \vskip\baselineskip
        \begin{subfigure}[b]{0.475\textwidth}   
            \centering 
            \includegraphics[scale=0.45, trim=90 210 50 220]{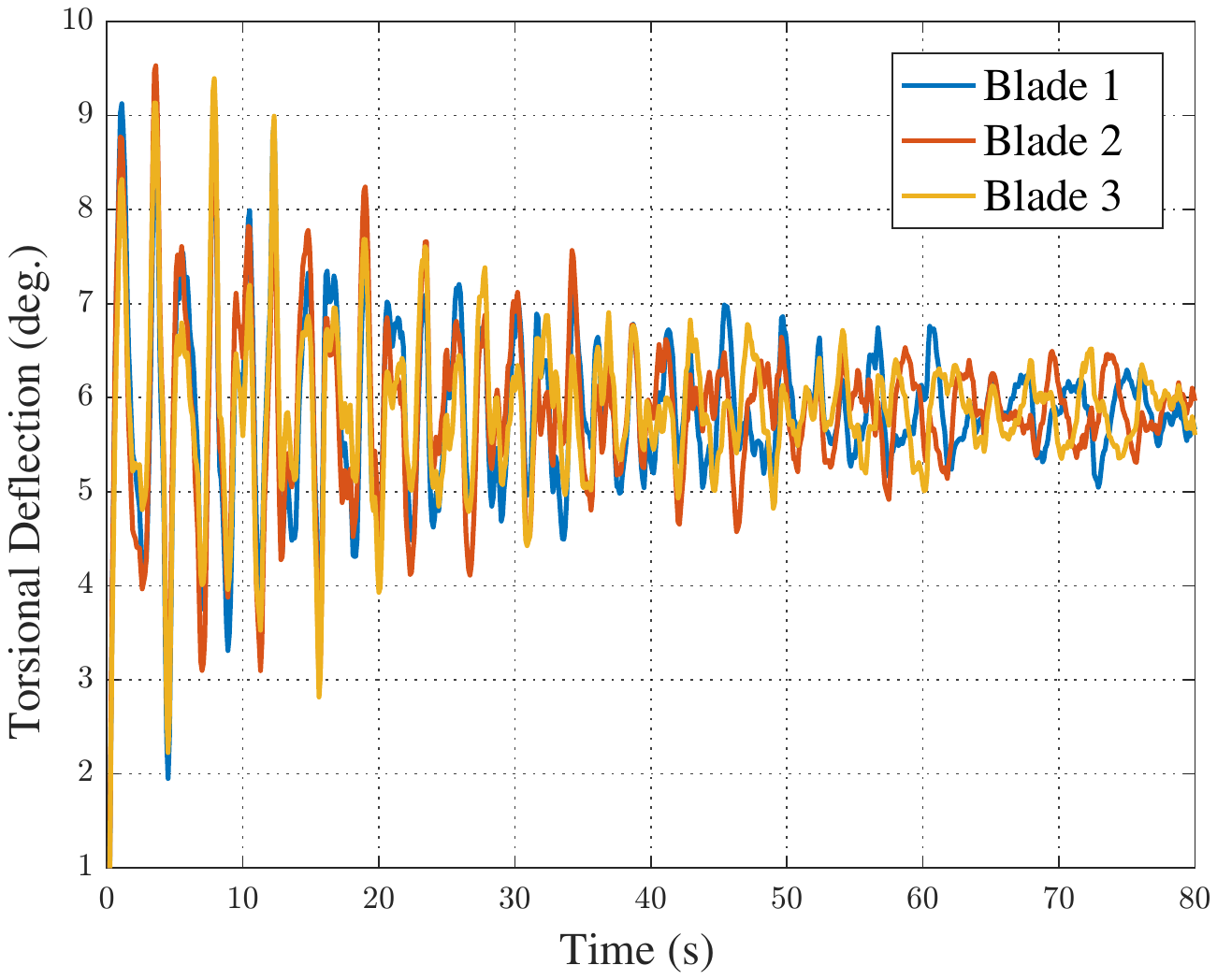}
            \caption[]%
            {{\small Torsional deflection}}    
            
        \end{subfigure}
        \caption[Time history of FOWT rotor-blade tip deflections at above-rated conditions]
        {Time history of rotor-blade tip deflections at above-rated operational conditions} 
        \label{AR_blade_deflect_fowts}
    \end{figure*}

Rotor-blade deflection behavior is shown in Fig.~\ref{AR_blade_deflect_fowts}. Though the amplitude of the flapwise deflections have been decreased due to the pitch of the rotor-blades, as compared to the rated operational case, the impact the rigid-body motions have had on the flapwise behavior is the greatest out of all of the operational conditions examined here. In the rated operation, the highest peak-to-peak about the mean deflection was about 1 m, while the above-rated rigid-body motion has loaded the blade 1.5 m peak-to-peak about the mean deflection. However, as it is for the below-rated and rated operation, the edgewise and torsional deflections are not appreciably impacted by the above-rated rigid-body motions.

\begin{figure*}[p!]
        \centering
        \begin{subfigure}[b]{0.475\textwidth}
            \centering
            \includegraphics[scale=0.45, trim=90 210 50 250]{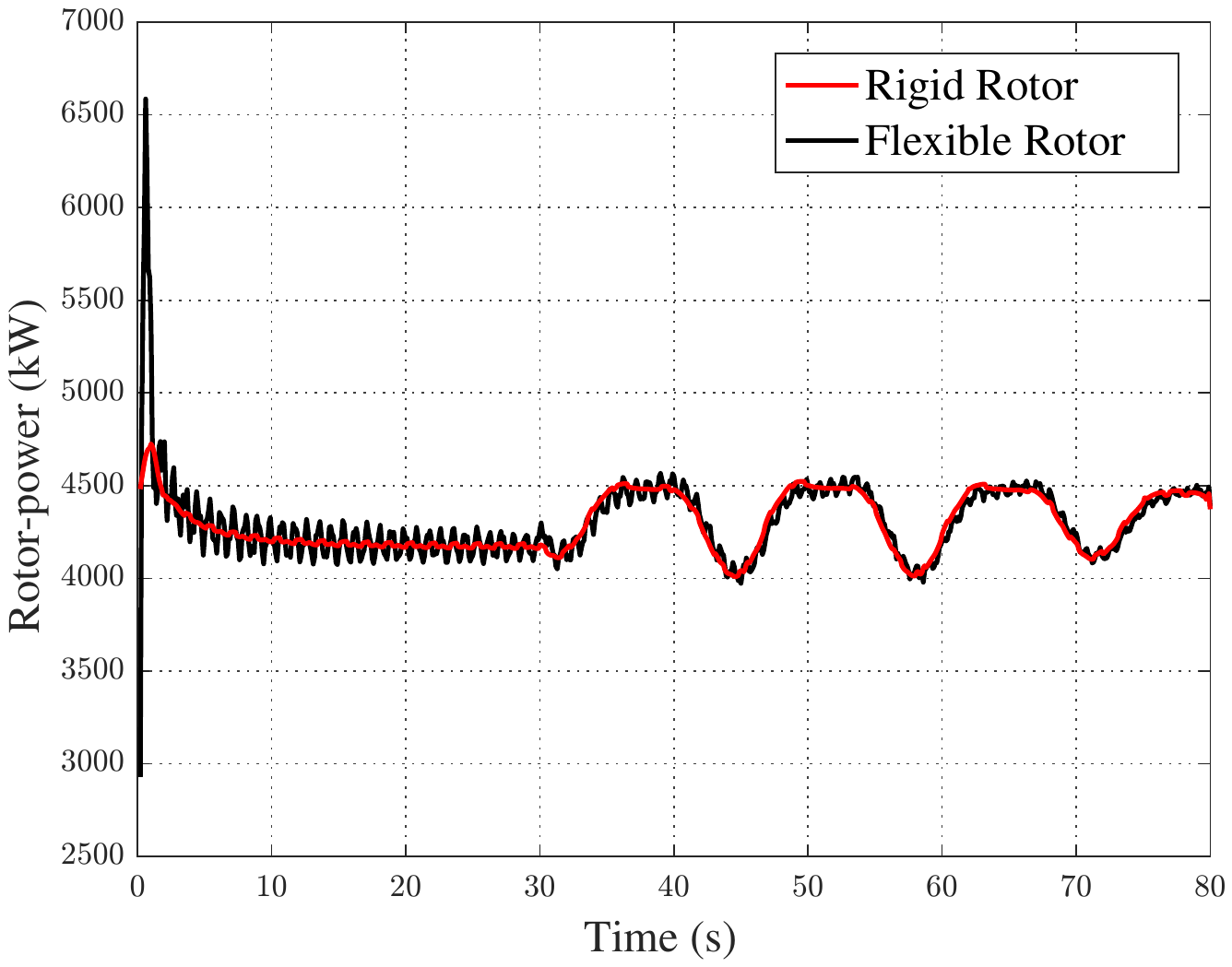}
            \caption[]%
            {{Rotor power}}    
            
        \end{subfigure}
        \hfill
        \begin{subfigure}[b]{0.475\textwidth}  
            \centering 
            \includegraphics[scale=0.45, trim=90 210 25 250]{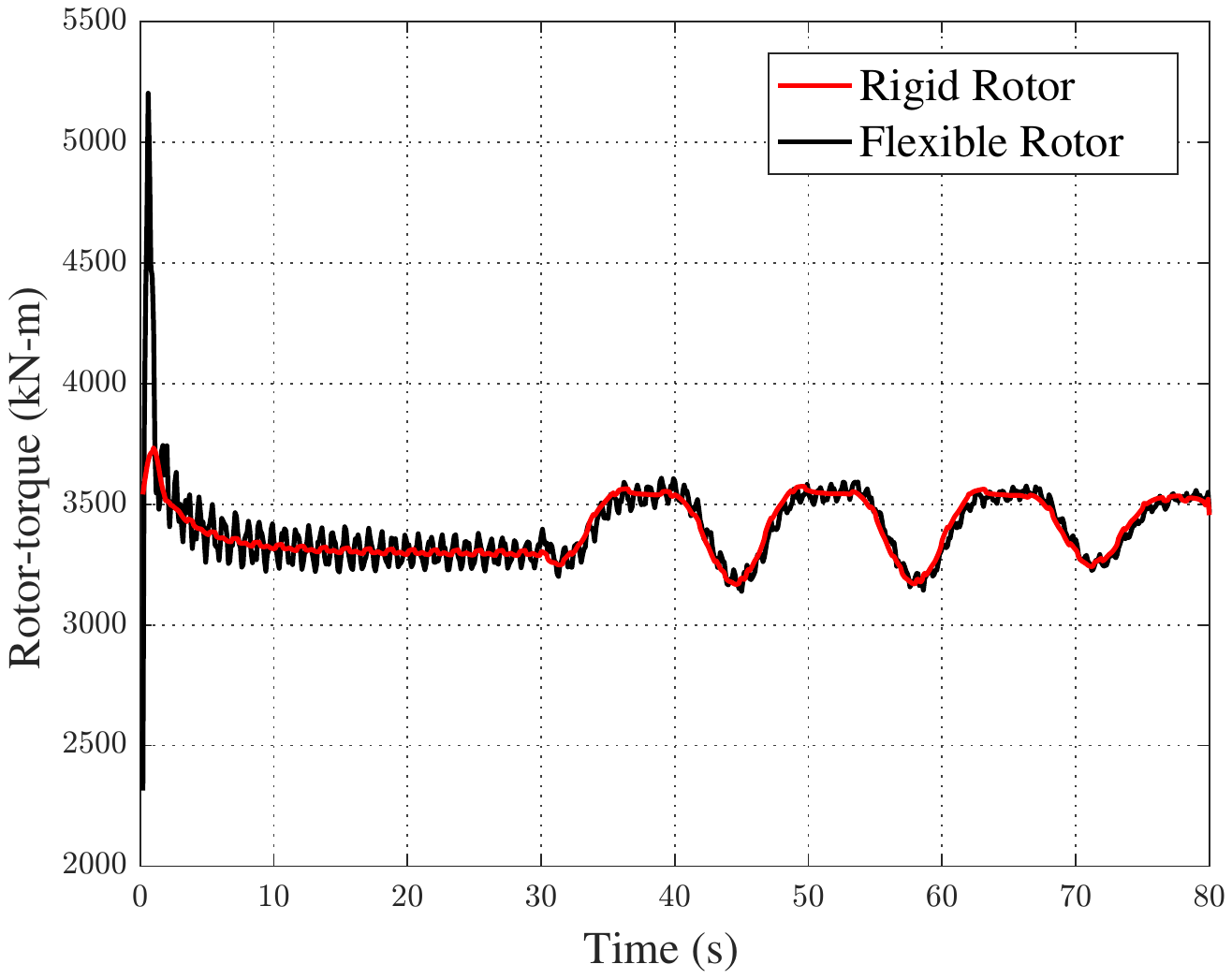}
            \caption[]%
            {{\small Rotor torque}}    
            
        \end{subfigure}
        \vskip\baselineskip
        \begin{subfigure}[b]{0.475\textwidth}   
            \centering 
            \includegraphics[scale=0.45, trim=90 210 50 220]{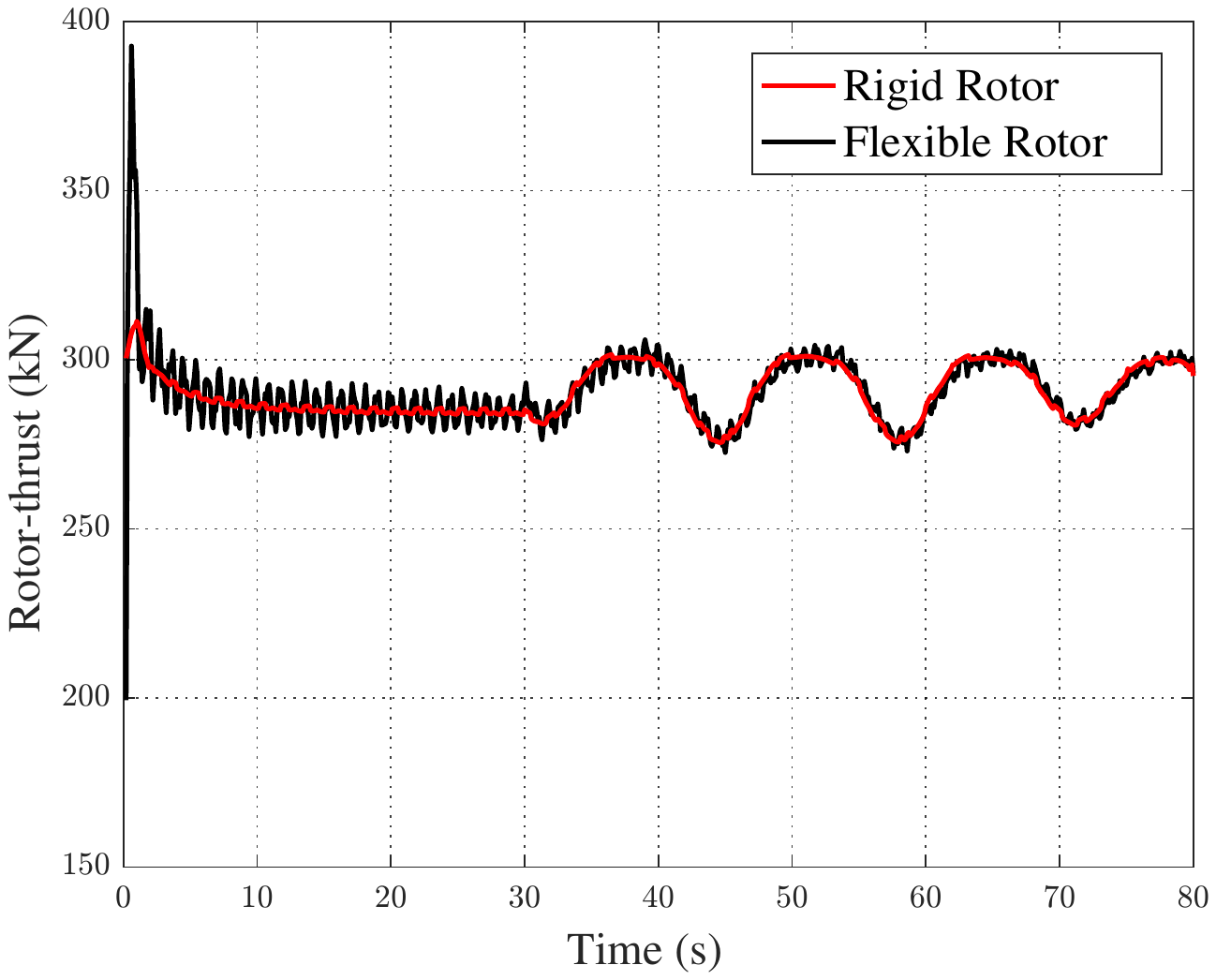}
            \caption[]%
            {{\small Rotor thrust}}    
            
        \end{subfigure}
        \caption[FOWT rotor performance metrics at above-rated operational conditions]
        {Rigid and flexible time history of rotor operational performance at above-rated operational conditions } 
        \label{AR_perf_metrics_fowts}
    \end{figure*}

The flexible rotor performance of the above-rated FOWT operation is demonstrated in Fig.~\ref{AR_perf_metrics_fowts}. The power, torque, and thrust also show peaks and troughs at the inflection points (zero instantaneous velocity) of the rigid-body motion presented in Fig.~\ref{AR_rbms}. The amplitudes of the performance fluctuations in power, torque, and thrust, are not as large as was seen in the rated case. This result is expected, as a relatively high blade pitch has been incorporated in the operation of the rotor-blades to reduce aerodynamic loads at above-rated wind speeds.

\subsection{Overview and Final Remarks of the FOWT Aeroelastic Results}

\begin{figure}[h!]
    \centering
    \includegraphics[scale=0.65, trim=0cm 8cm 0 8cm]{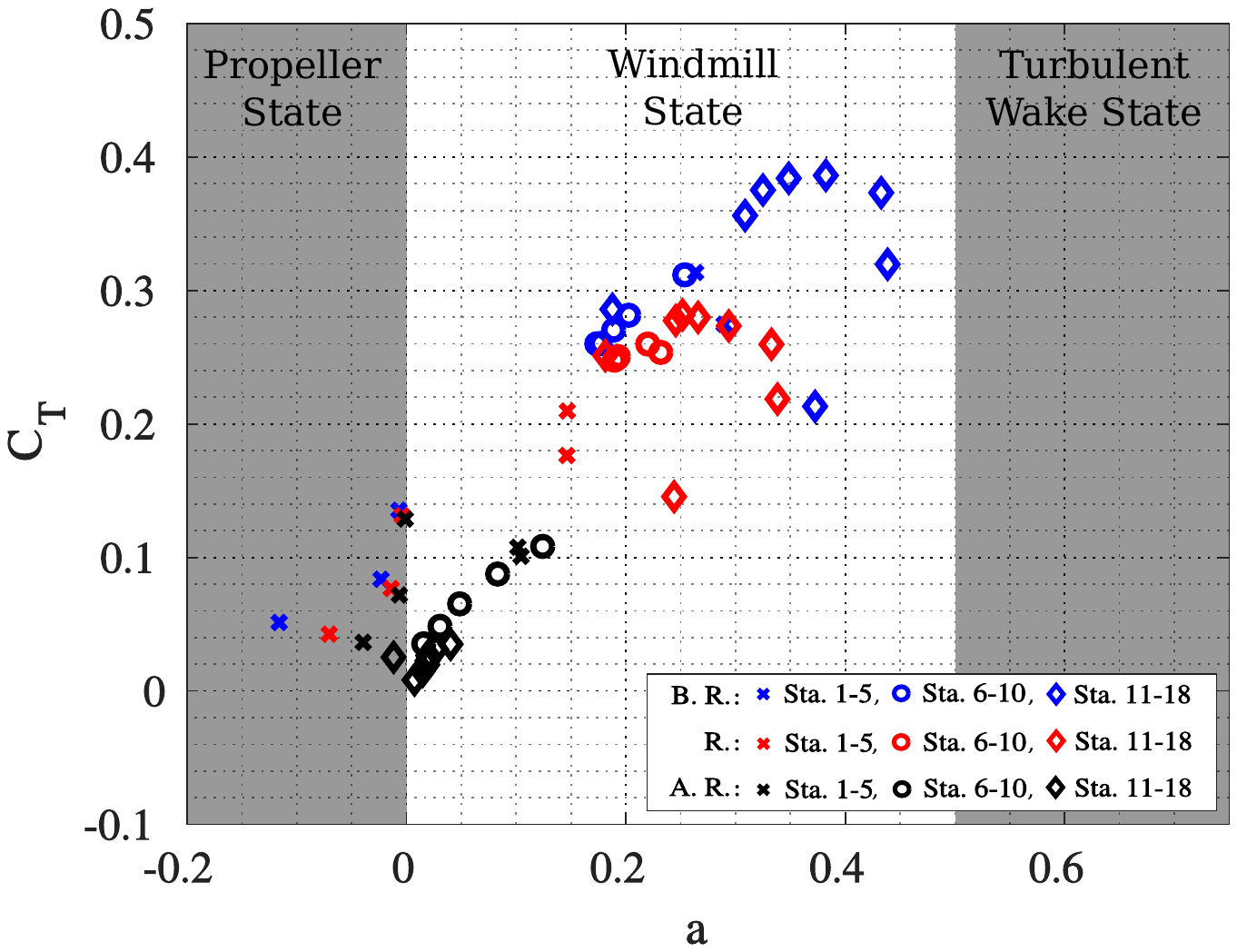}
    \caption{Rotor operational states at below-rated (B.~R.), rated (R.), and above-rated (A.~R.) conditions based on axial induction ($a$) and thrust coefficients ($C_T$) across spanwise stations (Sta.).}
    \label{avCT}
\end{figure}

The aeroelastic results of the floating offshore wind turbine at different operational conditions highlight very important dependencies between the rotor operation and its offshore environment. It is found that below-rated offshore environment of the OC3-Hywind spar buoy FOWT has no substantial impact on blade dynamics, rotor power, rotor torque, or rotor thrust. This result is to be expected due to the weak wave-induced motions of the below-rated offshore conditions. However, the rated offshore environment of the ITI-Energy barge FOWT had a notable impact on blade dynamics, where the blade vibrations oscillated with the frequency content of the wave-induced motions and rotational frequencies. Performance metrics such as rotor power, rotor torque, and rotor power were influenced by the wave-induced motion of the rated operational conditions. The performance metrics fluctuated with the frequency content of the wave-induced motions. The above-rated wave-induced motions generated the most impact on blade deflections about the mean, but not on the level of overall blade deflection magnitude of the rated operational case. However, performance metrics of the rotor operation at above-rated conditions were heavily impacted.  Finally, conclusions about the rotor operational states of the FOWT rotor at below-rated, rated, and above-rated operational conditions can be made by referring to Fig.~\ref{avCT}, i.e., at below-rated conditions most of the rotor is operating in the windmill state, where the blade tips operate near the turbulent wake state; at rated conditions the rotor is operating mostly in the windmill state; at the above-rated condition the rotor is operating near or in the propeller state.

\section{Conclusions}\label{Conclusions}
In the current Part 2 of this paper, the aeroelastic framework presented in Part 1 was employed to model below-rated, rated, and above-rated conditions of the NREL 5-MW reference floating offshore wind turbine rotor under pitching wave-induced motions. To the best of the authors' knowledge, the presented study is the first to model FOWTs under wave-induced pitching motions that use a strongly-coupled FVM-based aeroelastic framework. Due to a lack of experimental and numerical data to serve as benchmarks to test FOWT RWIs and corresponding aeroelastic phenomena, the current framework was tested against rotor performance results of the well-known WInDS aerodynamic tool, a moderate fidelity free-vortex wake method framework capable of modeling rigid rotor FOWT aerodynamics. Results of the current framework showed favorable agreement with WInDS with respect to mean values of rotor performance, i.e., rotor thrust, torque, and power. The current aeroelastic framework, however, has additional physical insight of rotor behavior than WInDS. Specifically, rotor blade deflections are captured with the current aeroelastic framework under the wave-induced pitching rotor motions caused by the offshore environment. It was found that below-rated wave-induced motions have a negligible impact on blade dynamics. However, rated and above-rated wave-induced motions impact the blade dynamics such that the blades oscillations contain a combined frequency content of the rotor rotation frequency and the wave-induced rotor motion. Therefore, the aeroelastic interaction between the blade deformations and the production of the rotor near-wake is pronounced in these cases.

The purpose of the present paper is to demonstrate and highlight the self-contained aeroelastic capabilities of the framework presented in Part 1 \cite{rodriguez_JRE_p1}, and to show the aeroelastic insight the framework is capable of providing, which can be a valuable tool in engineering design and analysis investigations, as the current framework provides moderate fidelity and an acceptable cost for engineer-level investigations of rotor-wake interactions and their corresponding aeroelastic phenomena. It is also important to note that the treatment of the inertial loading conditions due to rigid body motions, are rough approximations, where future work will improve more accurate representations of rotorblade dynamics, i.e.~gyroscopic and Coriolis forces, and the rigid-body dynamics generated by the offshore environment. Nevertheless, despite these approximate modeling, the aeroelastic results of FOWT operation presented in this paper are meant to establish a benchmark to assist future studies employing the current framework and/or developing improvements of an FVM-based aeroelastic framework. 

\section{Acknowledgements}

The authors wish to acknowledge the support of the Air Force Office of Scientific Research under awards FA9550-15-1-0148 and FA9550-19-1-0095 monitored by Drs.~Douglas Smith and Gregg Abate. The authors would also like to thank Dr.~Matthew Lackner, Dr.~Thomas Sebastian, and Evan Gaertner for their contributions to the rotor-aerodynamics research community, for opening WInDS to the public, and for insightful conversations and feedback about the current investigation over the years. 

%\section*{References}
%\bibliographystyle{unsrt}
\bibliography{mybibfile}

\end{document}